\documentclass[a4paper,fleqn,usenatbib]{mnras}

\usepackage[T1]{fontenc}
\usepackage{ae,aecompl}

\usepackage{graphicx}	
\usepackage{amsmath}	
\usepackage{amssymb}	
\usepackage{subfigure}
\usepackage{multirow}

\newcommand{\pcc}{\,{\rm cm}^{-3}}
\newcommand{\gcc}{\,{\rm g \, cm}^{-3}}

\newcommand{\um}{\, {\rm \mu m}}

\newcommand{\kel}{\, {\rm K}}

\newcommand{\msun}{\, {\rm M}_\odot}
\newcommand{\nh}{n_{\rm H}}

\newcommand{\pc}{\, {\rm pc}}

\newcommand{\myr}{\, {\rm Myr}}

\newcommand{\kms}{\, {\rm km \, s^{-1}}}
\newcommand{\av}{A_{\rm V}}

\title[Apparent filament widths]{Synthetic line and continuum observations of simulated turbulent clouds: the apparent widths of filaments}

\author[Priestley \& Whitworth]{
F. D. Priestley and A. P. Whitworth
\\
School of Physics and Astronomy, Cardiff University, Queen's Buildings, The Parade, Cardiff CF24 3AA, UK \\
}

\date{Accepted XXX. Received YYY; in original form ZZZ}

\pubyear{2020}

\begin{document}
\label{firstpage}
\pagerange{\pageref{firstpage}--\pageref{lastpage}}
\maketitle

\begin{abstract}
Filamentary structures are ubiquitous in observations of real molecular clouds, and also in simulations of turbulent, self-gravitating gas. However, making comparisons between observations and simulations is complicated by the difficulty of estimating volume-densities observationally. Here, we have post-processed hydrodynamical simulations of a turbulent isothermal molecular cloud, using a full time-dependent chemical network. We have then run radiative transfer models to obtain synthetic line and continuum intensities that can be compared directly with those observed. We find that filaments have a characteristic width of $\,\sim\!0.1 \pc$, both on maps of their true surface density, and on maps of their $850\um$ dust-continuum emission, in agreement with previous work. On maps of line emission from CO isotopologues, the apparent widths of filaments are typically several times larger because the line intensities are poorly correlated with the surface density. On maps of line emission from dense-gas tracers such as N$_2$H$^+$ and HCN, the apparent widths of filaments are $\la 0.1\pc$. Thus, current observations of molecular-line emission are compatible with the universal $0.1 \pc$ filament width inferred from {\it Herschel} observations, provided proper account is taken of abundance, optical-depth, and excitation considerations. We find evidence for $\sim 0.4 \kms$ radial velocity differences across filaments. These radial velocity differences might be a useful indicator of the mechanism by which a filament has formed or is forming, for example the turbulent cloud scenario modelled here, as against other mechanisms such as cloud-cloud collisions.
\end{abstract}


\section{Introduction}

Since the observation that far-infrared {\it Herschel} images of molecular clouds appear almost ubiquitously filamentary in nature \citep{andre2010}, from sub-parsec scales up to $\sim 100 \pc$ giant molecular filaments \citep{ragan2014}, most studies have accepted that this filamentary structure plays an important role in star formation, for example by controlling the accretion rate onto prestellar cores \citep{kirk2013,peretto2013} or altering the fragmentation properties of the gas \citep{sanchezmonge2014,ragan2015,beuther2015}. Filaments are commonly produced in simulations of turbulent clouds \citep{kirk2015,federrath2016} and cloud-cloud collisions \citep{balfour2015,matsuomoto2015}, or by Galactic shear forces \citep{duartecabral2017,smith2020}. How much, if anything, each of these mechanisms contributes to the formation of the the observed filaments is an open question, with important implications for our understanding of star formation, both within clouds and on a global scale.

Observationally, \citet{arzoumanian2011,arzoumanian2019} find that filaments identified in maps of dust-continuum intensity appear to have a characteristic width of $\sim 0.1 \pc$, implying that any theoretical or numerical model ought to reproduce this in order to be considered valid. \citet{kirk2015} and \citet{federrath2016} both find that simulations of turbulent, self-gravitating clouds produce filaments of this approximate width for typical interstellar medium (ISM) conditions, irrespective of whether magnetic fields are included, but with a somewhat larger range of widths than found by \citet{arzoumanian2011,arzoumanian2019}. However, for simulated filaments the surface density is known exactly, whereas the observationally inferred surface densities depend on several assumptions about dust properties such as dust temperature, dust optical properties and gas-to-dust ratio, all of which are poorly constrained and may vary within the filament (e.g. \citealt{ossenkopf1994,howard2019}). If molecular line radiation is used to define filaments (rather than dust continuum emission), the inferred filament widths depend on the molecule oberved, from $\sim 0.01 \pc$ for N$_2$H$^+$ \citep{hacar2018} to $0.4 \pc$ for $^{13}$CO \citep{panopoulou2014}, although \citet{suri2019} find a value in agreement with \citet{arzoumanian2011} using C$^{18}$O. \citet{smith2014} and \citet{panopoulou2017} have also raised concerns about the effect of the fitting method on estimates of filament width. It is therefore unclear whether the correspondence between observed and simulated filament widths is really showing that the simulations are correct.

The above issues can be mitigated by converting the output of hydrodynamical simulations into synthetic observations of commonly detected molecules. These can then be compared directly with observed filaments. Previous work has largely focused on emission from [C II] and CO isotopologues (e.g. \citealt{penaloza2018, franeck2018}), the abundances of which are already calculated in many simulations in order to determine cooling rates, on scales from giant molecular clouds \citep{duartecabral2016} to individual star-forming filaments \citep{clarke2018}. However, the chemical networks used (e.g. \citealt{glover2012}) often neglect freeze-out of CO onto dust grains, which can significantly reduce its abundance in the densest regions; within these regions even the rarest isotopologues may also be optically thick. \citet{smith2012,smith2013} investigate emission from other molecular species, but assume an abundance for each species, rather than calculating it self-consistently. In this paper, we track the full time-dependent chemical evolution of a turbulent filamentary cloud, allowing us to generate synthetic line emission maps for any molecule in our chemical network. By computing the emergent intensities of many different molecular lines, tracing a range of gas densities, and then comparing them with the observed intensites, we are able to significantly improve constraints on the structure of star forming filaments.

\section{Method}

We model the hydrodynamical evolution of a turbulent molecular cloud using the Smoothed Particle Hydrodynamics (SPH) code {\sc phantom} \citep{price2018}. We consider a uniform density spherical cloud of mass $M = 200 \msun$ and radius $R = 1 \pc$, giving an initial density of $3.23 \times 10^{-21} \gcc\;(\equiv 680\,{\rm H_{_2}\,cm^{-3}})$. The gas in the cloud is isothermal at $T=10\kel$  and the hydrogen is molecular, hence the isothermal sound speed is $c_{\rm s} = 0.2 \kms$. The cloud is at the centre of a cubic box with side length $4R$. Outside the cloud the density is $3.23 \times 10^{-23} \gcc\;(\equiv 6.8\,{\rm H_{_2}\,cm^{-3}})$, the gas is isothermal at $T=1000 \kel$, and the hydrogen is molecular, hence the isothermal sound speed is $c_{\rm s} = 2 \kms$. There is therefore  pressure balance across the boundary of the cloud. A turbulent velocity field is generated with a power spectrum $P(k) \propto k^{-4}$, wave-numbers between $k_{\rm min}=1$ and $k_{\rm max}=100$ (where $k_{\rm min}=1$ corresponds to wavelength $\lambda =R$), and a thermal mix of compressive and solenoidal modes. The SPH-particle velocities are scaled so that the initial root-mean-squared Mach number is $5$. These values give similar turbulent velocity fields to those simulated by \citet{kirk2015} and \citet{federrath2016}, although some details, such as the implementation of turbulence, differ. {We use $\sim 140\,000$ particles in the cloud, and a further $\sim 20\,000$ outside the cloud, with a particle mass resolution of $1.4 \times 10^{-3} \msun$. The mean number of neighbours within an SPH smoothing kernel is 57. Increasing the mass resolution by a factor of 5, by increasing the number of SPH particles to $\sim 800,000$, does not noticeably change the results.}

For isothermal gas where hydrogen is already in molecular form, the chemical evolution is almost completely decoupled from the hydrodynamics. We can therefore post-process our SPH results with a time-dependent chemical code, as in \citet{priestley2019}. We use a subset of $10\,000$ SPH particles,\footnote{This means that even SPH particles whose chemical evolution is not followed are likely to have several (on average 6) SPH particles within their smoothing kernel whose chemical evolution {\it is} followed.} chosen randomly from those with initial positions inside the cloud, and input their density evolution as a function of time into {\sc ucl\_chem} \citep{holdship2017}, which uses the UMIST12 reaction network \citep{mcelroy2013} with additional molecular freeze-out and grain surface reactions as described in \citet{holdship2017}. For all SPH particles, we assume a constant gas-kinetic temperature ($T=10 \kel$ for those in the cloud, and $T = 1000 \kel$ for those outside the cloud), a dust temperature $T_{\rm dust} = 10 \kel$, a cosmic ray ionization rate {per H$_2$ molecule} of $1.3 \times 10^{-17} \, {\rm s^{-1}}$, and elemental abundances from \citet{lee1998}.

\begin{figure*}
  \centering
  \subfigure{\includegraphics[width=\columnwidth]{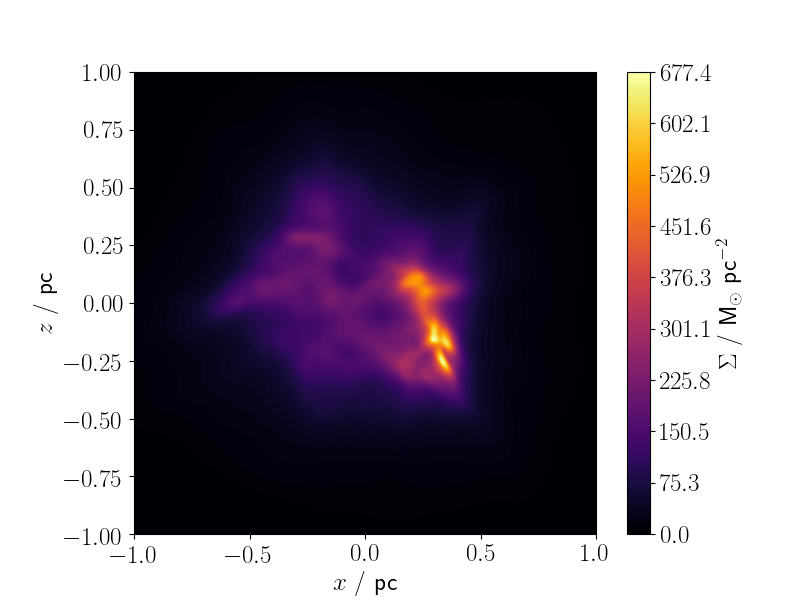}}\quad
  \subfigure{\includegraphics[width=\columnwidth]{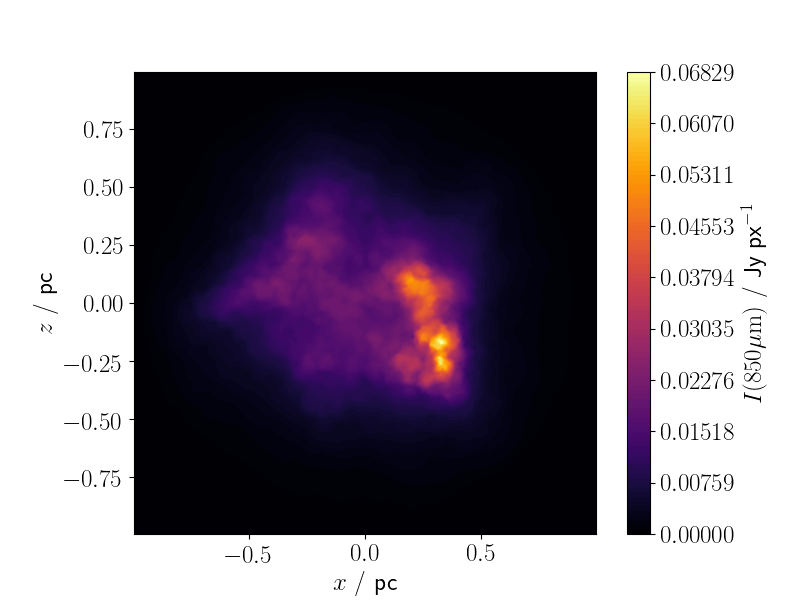}}\\
  \subfigure{\includegraphics[width=\columnwidth]{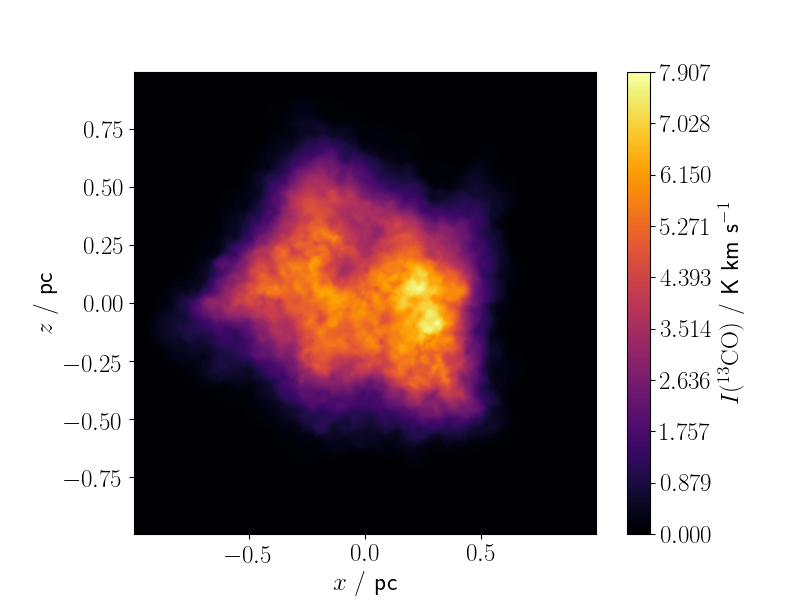}}\quad
  \subfigure{\includegraphics[width=\columnwidth]{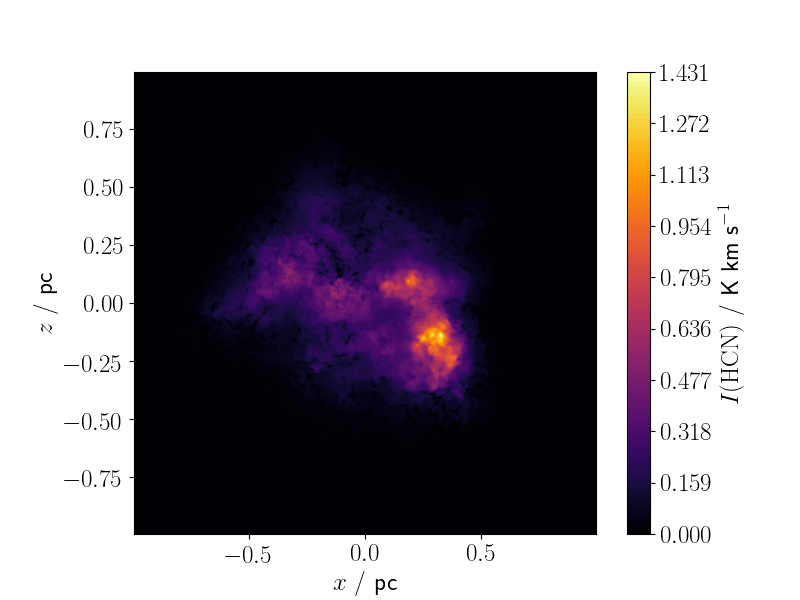}}
  \caption{Images projected on the $y\!=\!0$ plane for the $G\!=\!1$ case at $0.868\,{\rm Myr}$. {\it Top left:} surface density. {\it Top right:} monochromatic intensity of dust-continuum emission at $850 \um$. {\it Bottom left:} integrated intensity of the ${\rm ^{13}CO}\,(J\!=\!1-0)$ line. {\it Bottom right:} integrated intensity of the HCN$\,(J\!=\!1-0)$ line.}
  \label{fig:coldens}
\end{figure*}

In some previous work on prestellar cores, the visual extinction $\av$ at each point in the cloud has either been calculated exactly exploiting imposed symmetries (e.g. \citealt{aikawa2005,tassis2012,priestley2018}), or has been assumed to be large enough, due to external shielding, to make the radiation field negligible \citep{priestley2019}. The former situation does not apply to turbulent clouds, while on scales of $\,\gtrsim\!1 \pc$ the second is not justifiable either. Several algorithms have been developed to estimate the shielding column density (and hence $\av$) in hydrodynamical simulations (e.g. \citealt{clark2012}). However, these still entail significant computational expense to obtain useful resolution. Here we approximate $\av$ using a modified version of the method of \citet{dobbs2008}, who assumed that the effective column density is just the local density multiplied by a length scale. While those authors used a constant length scale, we instead use the local Jeans length, defined as $c_{\rm s}(G\rho)^{-1/2}$, which we find returns column densities comparable to the maximum midplane value for the densest gas. We convert the column density to $\av$ with a typical ISM factor of $6.0\times 10^{-22}\,{\rm mag\,(\rm H\,cm^{-2}})^{-1}$ \citep{bohlin1978}. We investigate ambient far-ultraviolet (FUV) radiation field strengths of $G = 0,1$ and $5$ measured in Habing units \citep{habing1968}. $G\!=\!1$ is the fiducial case, on which we focus the discussion.

\begin{figure*}
  \centering
  \subfigure{\includegraphics[width=\columnwidth]{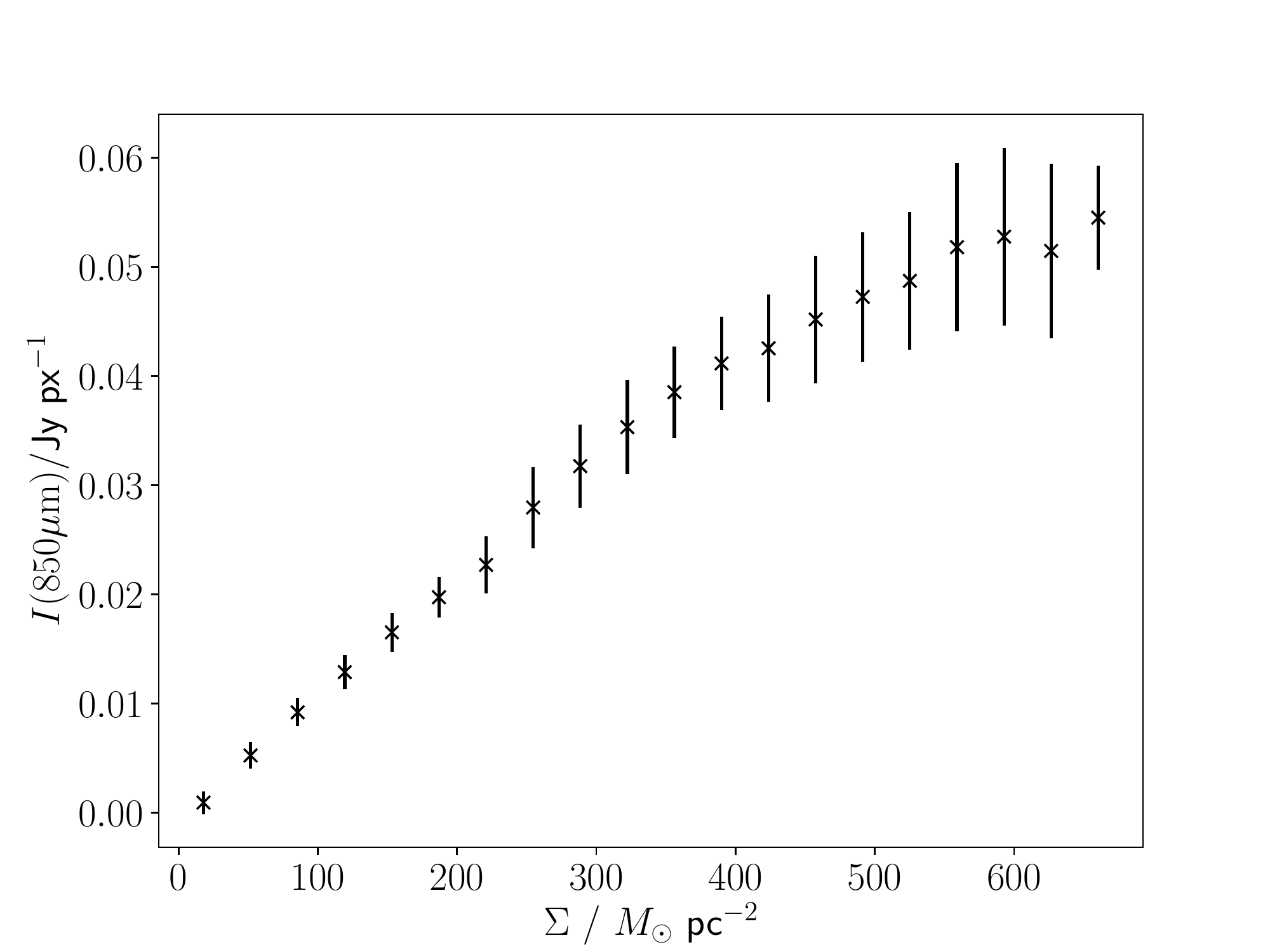}}\quad
  \subfigure{\includegraphics[width=\columnwidth]{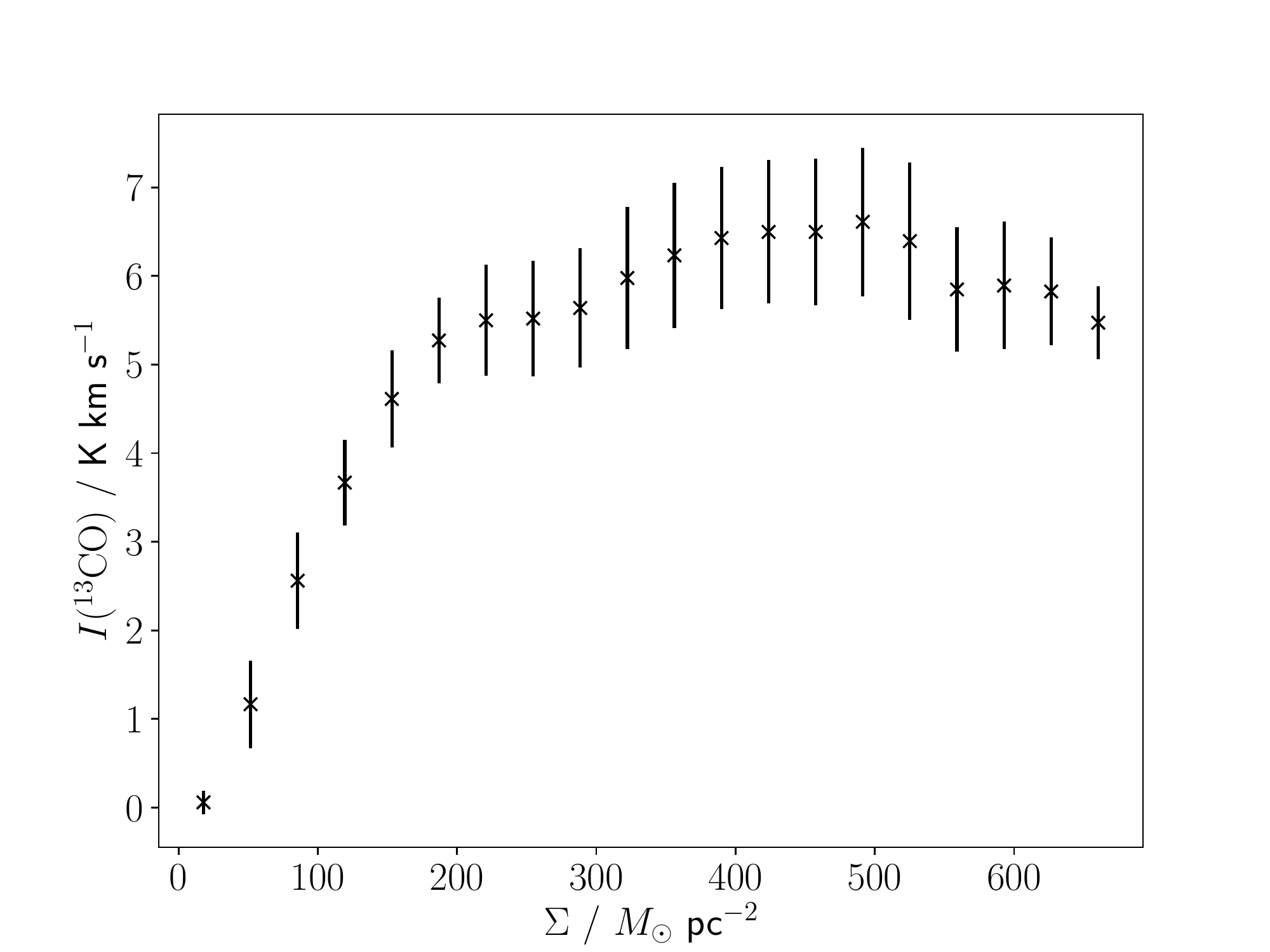}}
  \caption{The ratio of intensity to surface density on the $y\!=\!0$ plane for the $G\!=\!1$ case at $0.868\,{\rm Myr}$. {\it Left:} monochromatic intensity of dust-continuum emission at $850 \um$. {\it Right:} integrated line intensity of ${\rm ^{13}CO}\,J\!=\!1-0$ emission. The points give mean values, and the error bars give standard deviations.}
  \label{fig:alpha}
\end{figure*}

\begin{figure*}
  \centering
  \subfigure{\includegraphics[width=\columnwidth]{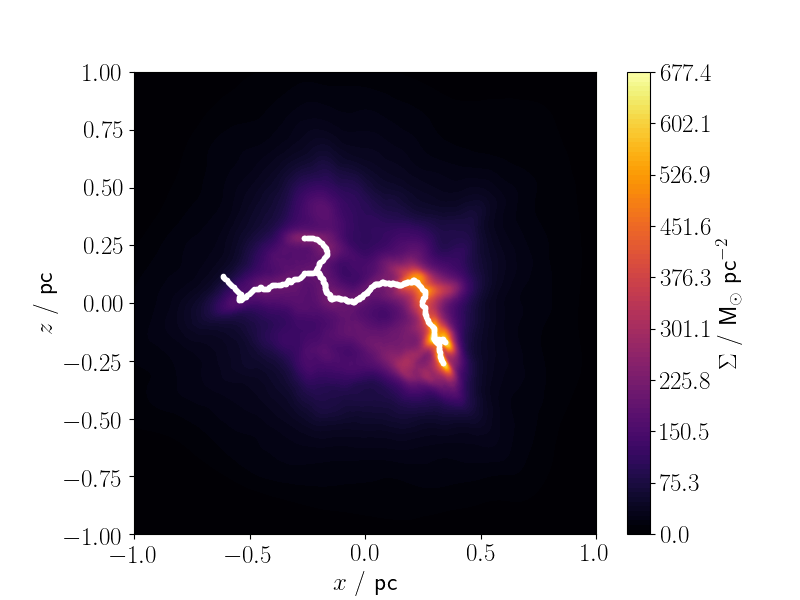}}\quad
  \subfigure{\includegraphics[width=\columnwidth]{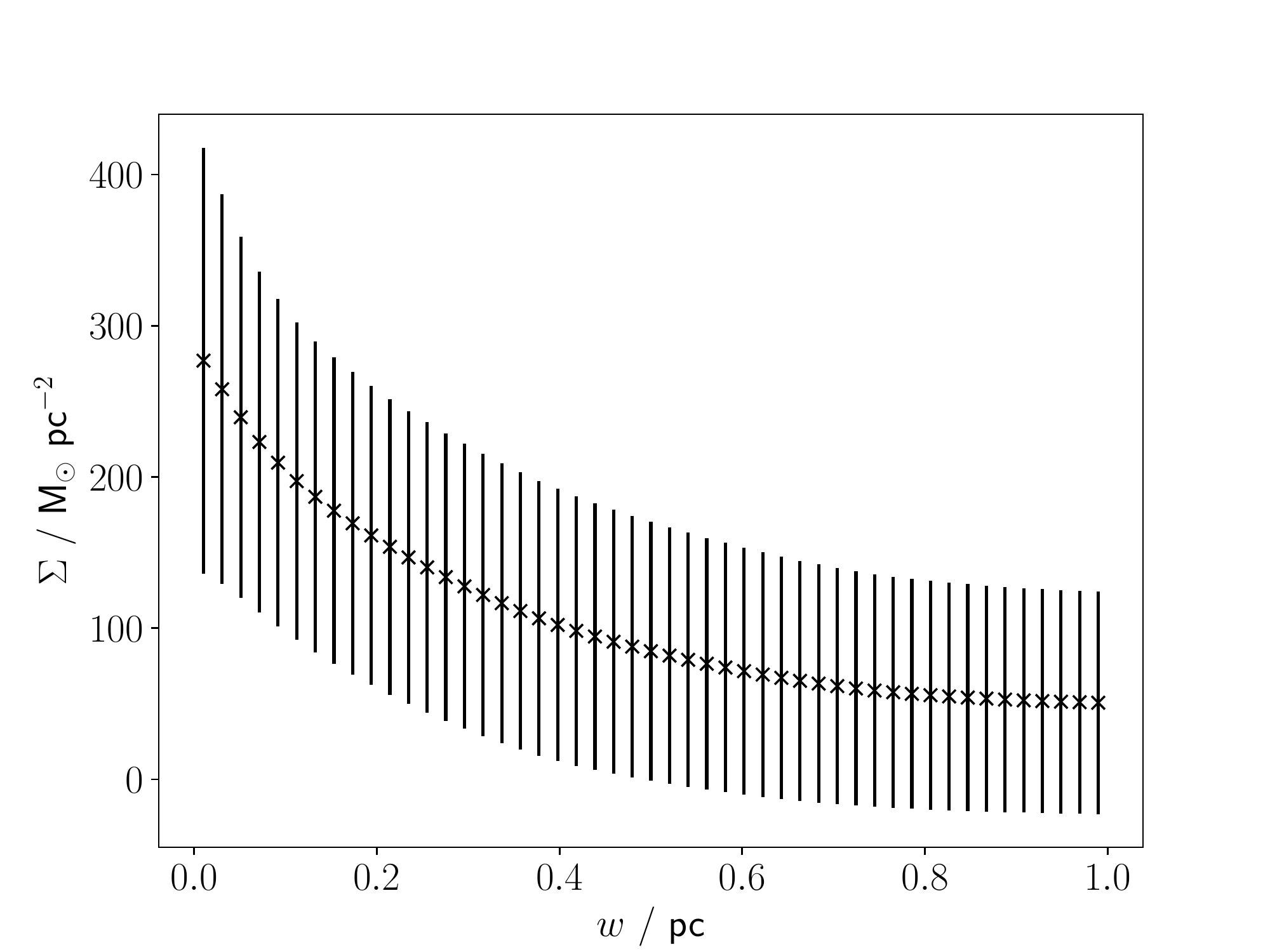}}
  \caption{The $G\!=\!1$ case at $0.868\,{\rm Myr}$, projected on the $y\!=\!0$ plane. {\it Left:} surface density image with white pixels tracing the filament spine as identified by DisPerSE. {\it Right:}  average surface-density profile, $\Sigma(w)$, perpendicular to the filament; the points give mean values, and the error bars give standard deviations.}
  \label{fig:filaments}
\end{figure*}

Our chemical modelling gives us the detailed three-dimensional chemical composition of the cloud at each point in time during the simulation. We use {\sc lime} \citep{brinch2010} to convert this into synthetic dust-continuum and molecular-line observations, using dust optical properties from \citet{ossenkopf1994} and molecular data from the {\sc lamda} database \citep{schoier2005}. { We use $10,000$ randomly-distributed sample points, assigning each point the density, molecular abundance and velocity of the nearest SPH particle. While this undersamples denser regions, compared to the distribution of SPH particles, we find that increasing the resolution does not substantially change our results.} We generate images in dust continuum emission, and in various molecular lines, as observed along the coordinate axes (i.e. projected on the $x\!=\!0$, $y\!=\!0$ and $z\!=\!0$ planes), and compare these with the true surface-density, e.g. for images projected on the $y\!=\!0$ plane
\begin{eqnarray}
\Sigma(x,z)=\int\limits_{y=-2R}^{y=+2R}\rho(x,y,z)\,dy\,.
\end{eqnarray}
{ We use a pixel size of $0.005 \pc$, comparable to the typical SPH smoothing length.} We identify filaments using the DisPerSE algorithm \citep{sousbie2011}. The persistence threshold required by DisPerSE is computed from the intensities outside a projected radius of $R$, which do not contain any cloud particles and therefore represent a `background' noise level. Our chemical network does not distinguish between isotopes of the same element and ignores chemical fractionation; when calculating the emission from isotopologues, we assume ${\rm ^{12}C/^{13}C}=100$ and ${\rm ^{16}O/^{18}O}=500$.

\section{Results}

Figure \ref{fig:coldens} shows images on the $y\!=\!0$ plane of the surface density, the monochromatic intensity of dust-continuum emission at $850\um$, and the integrated intensities of the $J\!=\!1-0$ lines from $^{13}$CO and HCN, for the $G=1$ case after $0.868 \myr$, { at which point the range of surface densities is comparable to the models presented in \citet{federrath2016}.} One can identify by eye filamentary structures in the surface-density image, and the dust continuum emission also traces the surface density accurately. There is much lower contrast between lower and higher surface-density gas in the $^{13}$CO line emission, and consequently this emission map appears more lumpy than filamentary. The HCN emission is limited to the highest density regions.

This can be seen more clearly in the ratios of intensity to surface density, shown in Figure \ref{fig:alpha}. The $850 \um$ intensity increases approximately in proportion to the surface density up to $\Sigma\sim 500 \msun \pc^{-2}$, but above this value the $850 \um$ intensity is somewhat underestimated. This is because the surface-density maps are evaluated using {\sc splash} \citep{price2007}, which uses all the SPH particles and an accurate interpolation procedure; in contrast, the $850\um$ dust emission maps produced by {\sc lime} have lower resolution, because (a) they only use the chemical tracer particles, and (b) {\sc lime} adopts a much cruder interpolation procedure to obtain the dust density along the line of sight. The $^{13}$CO $J\!=\!1-0$ integrated intensity `saturates' at $\Sigma\sim 250 \msun \pc^{-2}$, because it becomes optically thick, and for $\Sigma\ga 500 \msun \pc^{-2}$ it declines due to freeze-out of $^{13}$CO. Despite this lack of sensitivity to high densities, the filamentary spines traced by DisPerSE for both the dust $850\um$ and the $^{13}$CO $J\!=\!1-0$ intensity maps, are very similar to the actual surface-density map. For consistency, we use the filament spine traced using the surface-density map (as shown in the left panel of Figure \ref{fig:filaments}) to determine filament properties in all observational tracers.

\begin{figure}
  \centering
  \includegraphics[width=\columnwidth]{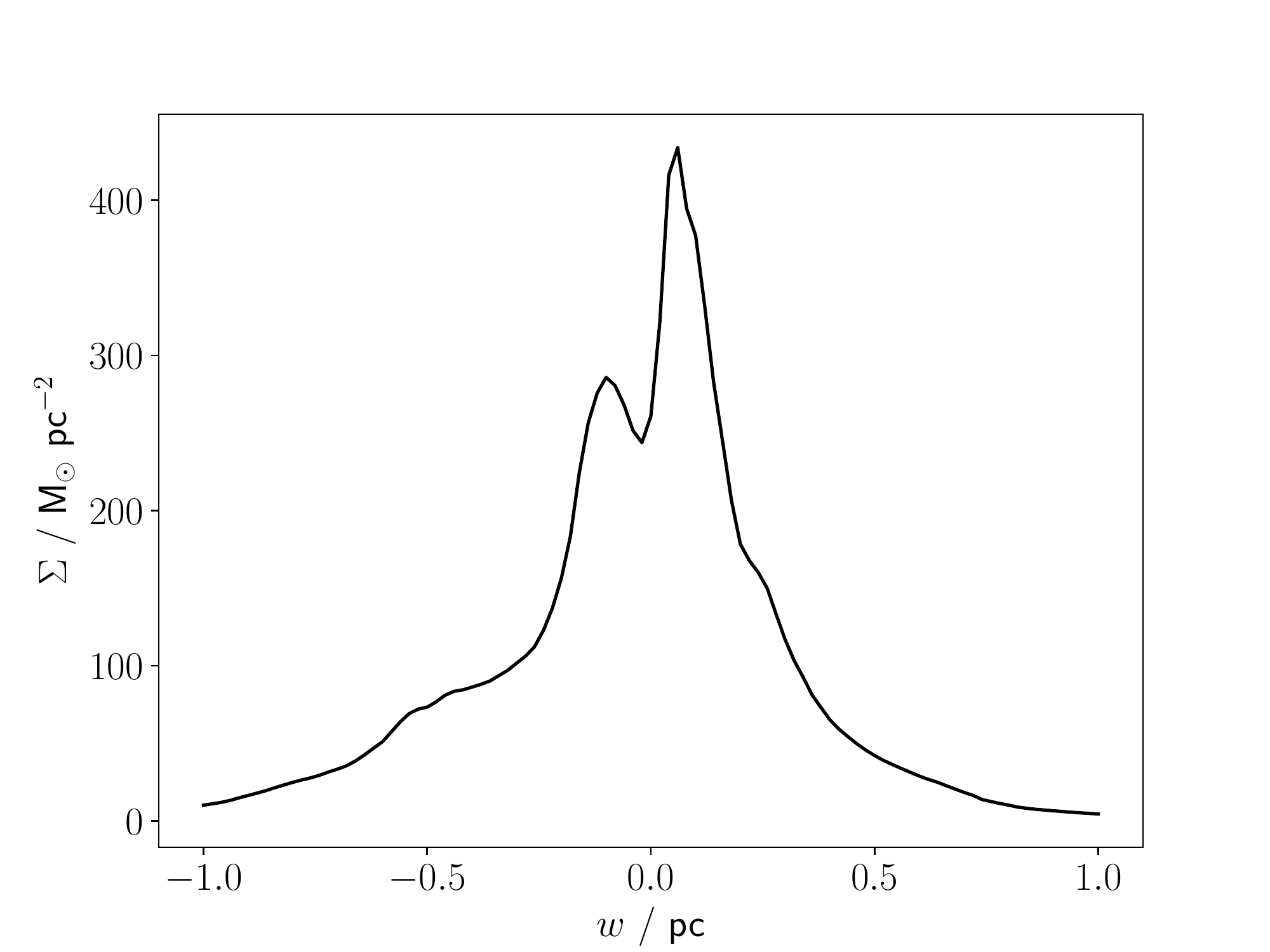}
  \caption{Average surface density profile perpendicular to the filamentary structure traced by DisPerSE from the surface-density map on the $y\!=\!0$ plane for the $G\!\!=1$ case at $0.868\,{\rm Myr}$.}
  \label{fig:slice}
\end{figure}

The right panel of Figure \ref{fig:filaments} shows the mean surface-density profile perpendicular to the filament spine, $\bar{\Sigma}_\perp\!(w)$, where $w$ is the impact parameter relative to the spine, and the averaging is over all the pixels defining the spine. To calculate this profile, we identify for each general pixel, $p$, the closest spine pixel, $s$, and compute the distance between them, $w_{sp}$. The surface-density, $\Sigma_p$, at general pixel $p$ then contributes to the surface-density profile perpendicular to the filament at spine pixel $s$, i.e. $\Sigma_{\perp s}(w_{sp})=\Sigma_p$. The mean surface-density profile $\bar{\Sigma}_\perp\!(w)$ is obtained by averaging $\Sigma_{\perp s}(w)$ over all spine pixels, $s$. Whereas \citet{federrath2016} find sharply peaked profiles with a well-defined typical Full Width at Half Maximum (FWHM) of $\sim 0.1 \pc$, our profiles are much broader, with mean FWHM$\,\sim 0.6 \pc$, and significant scatter around the mean at all impact parameters. This is a consequence of the averaging process. For an individual filament spine pixel, as identified by DisPerSE, the surface-density profile may have multiple, off-centre peaks, as illustrated by Figure \ref{fig:slice}. These individual peaks have widths comparable to those seen in \citet{federrath2016}, but the effect of averaging all points together results in a much broader distribution. A very similar phenomenon has been found by \citet{suri2019} in C$^{18}$O observations of the Orion A molecular cloud, who show that the width of the average filament profile is much broader than the typical width of the individual density peaks.

We note that, although many of our simulation parameters are very similar to those of \citet{federrath2016}, there are some significant differences: our simulation only includes an initial turbulent velocity field, which is then allowed to decay, as opposed to continuously driven turbulence; and we consider an isolated cloud, rather than a periodic box.

\begin{table*}
  \centering
  \caption{Filament HWHM median values and $16^{\rm th}$ and $84^{\rm th}$ percentiles after $0.868 \myr$, in $\pc$, for differing radiation field strengths, $G$, based on surface density, dust continuum and various molecular line intensities.}
  \begin{tabular}{cccccccccc}
    \hline
    $\;\;G\;\;$ & & $\Sigma$ & $I({\rm 850 \um})$ & $^{13}$CO & C$^{18}$O & NH$_3$ & HCN  & HCO$^+$  & N$_2$H$^+$  \\
    \hline
    \multirow{2}{*}{$0$} & $\;\;$median$\;\;$ & $0.16$ & $0.18$ & $0.34$ & $0.26$ & $0.18$ & $0.10$ & $0.12$ & $0.12$ \\
    & range & $0.08-0.40$ & $0.10-0.40$ & $0.22-0.56$ & $0.16-0.48$ & $0.12-0.32$ & $0.06-0.16$ & $0.08-0.22$ & $0.06-0.22$ \\
    \multirow{2}{*}{$1$} & median & $0.16$ & $0.18$ & $0.30$ & $0.24$ & $0.12$ & $0.12$ & $0.10$ & $0.06$ \\
    & range & $0.08-0.40$ & $0.10-0.40$ & $0.20-0.52$ & $0.14-0.44$ & $0.08-0.20$ & $0.06-0.20$ & $0.06-0.20$ & $0.04-0.12$ \\
    \multirow{2}{*}{$5$} & median & $0.16$ & $0.18$ & $0.26$ & $0.20$ & $0.10$ & $0.10$ & $0.08$ & $0.06$ \\
    & range & $0.08-0.40$ & $0.10-0.40$ & $0.16-0.44$ & $0.12-0.38$ & $0.06-0.18$ & $0.06-0.18$ & $0.06-0.14$ & $0.04-0.10$ \\
    \hline
  \end{tabular}
  \label{tab:widths}
\end{table*}

\begin{figure*}
  \centering
  \subfigure{\includegraphics[width=\columnwidth]{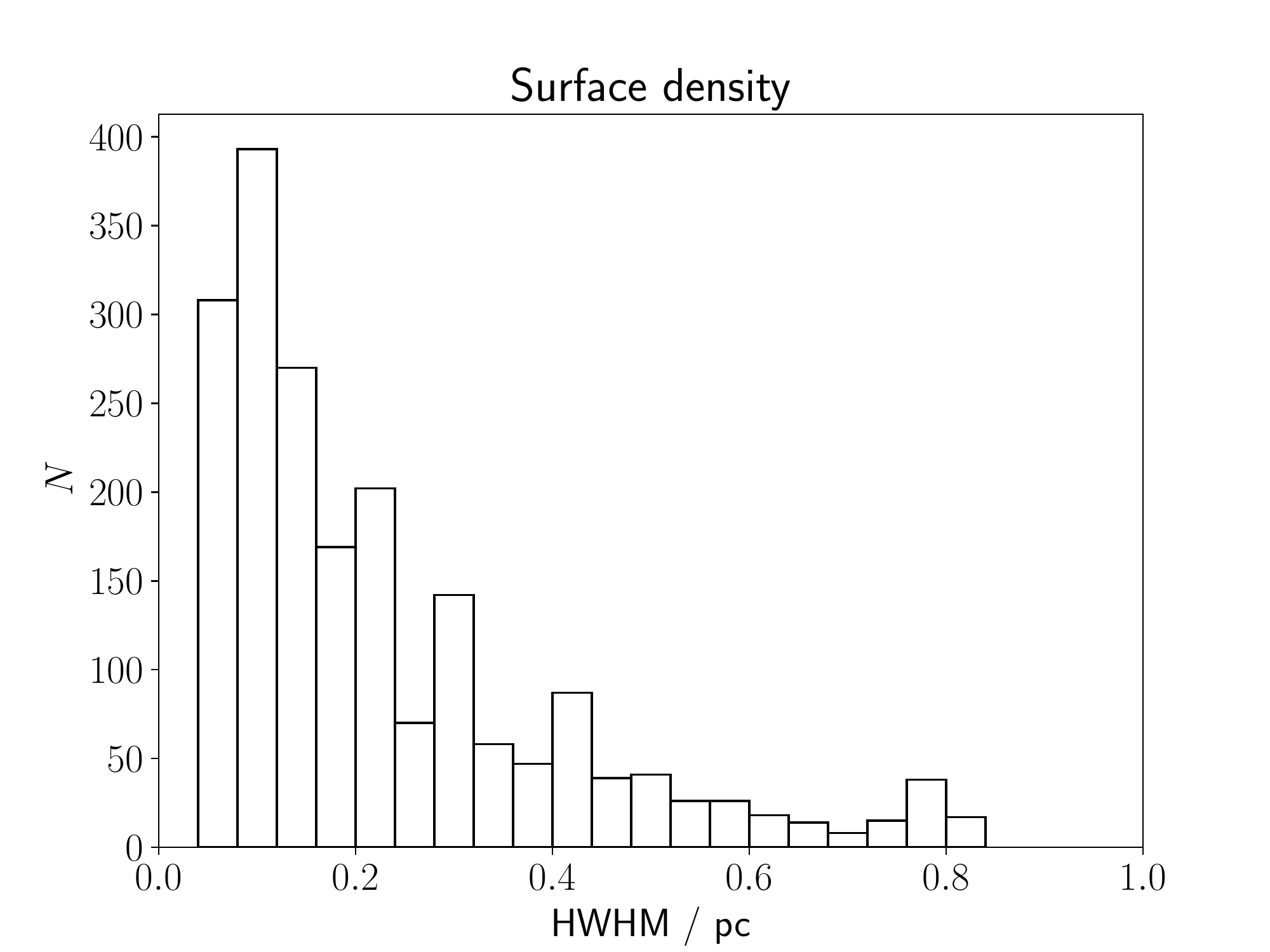}}\quad
  \subfigure{\includegraphics[width=\columnwidth]{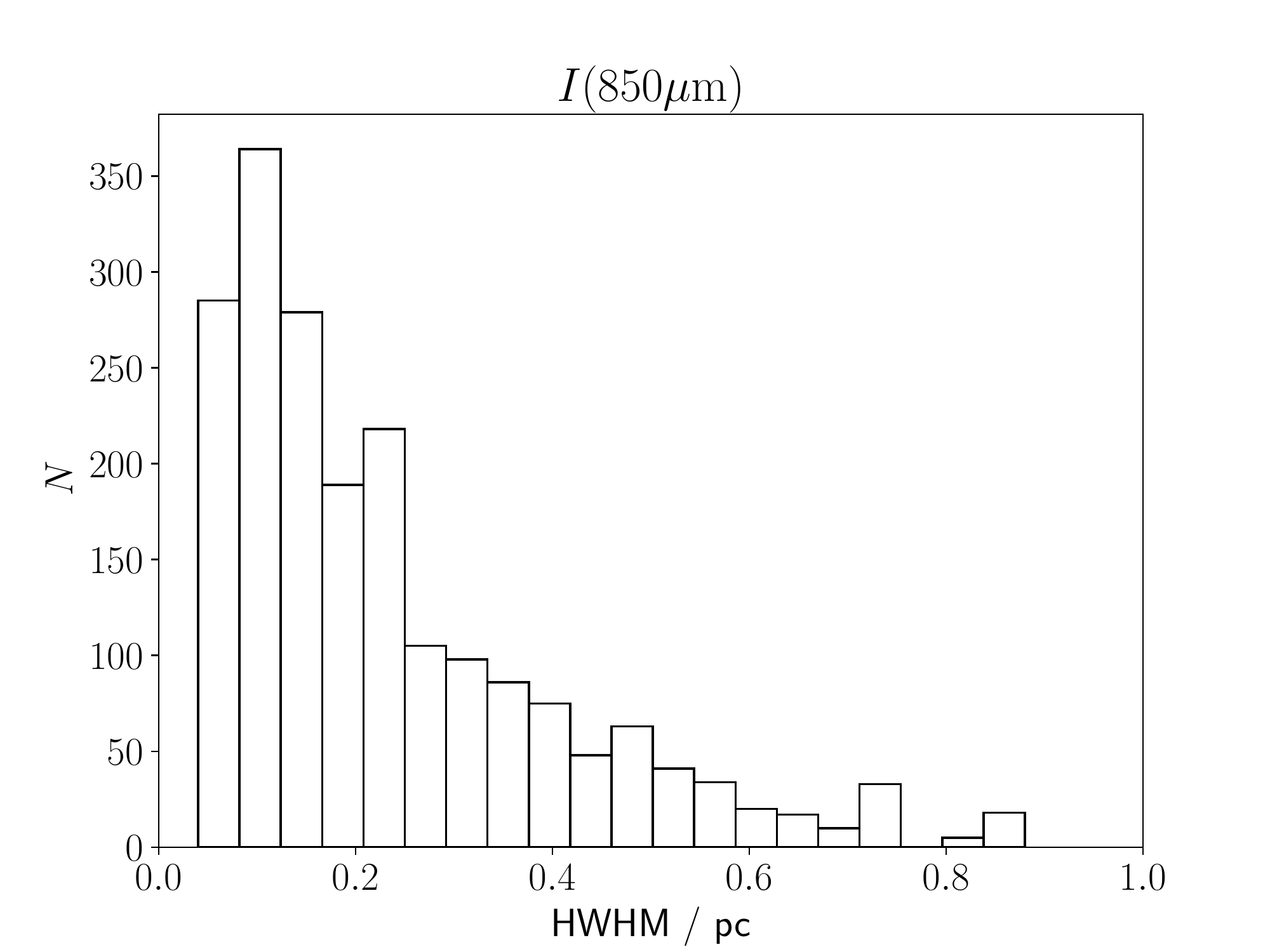}}\\
  \subfigure{\includegraphics[width=\columnwidth]{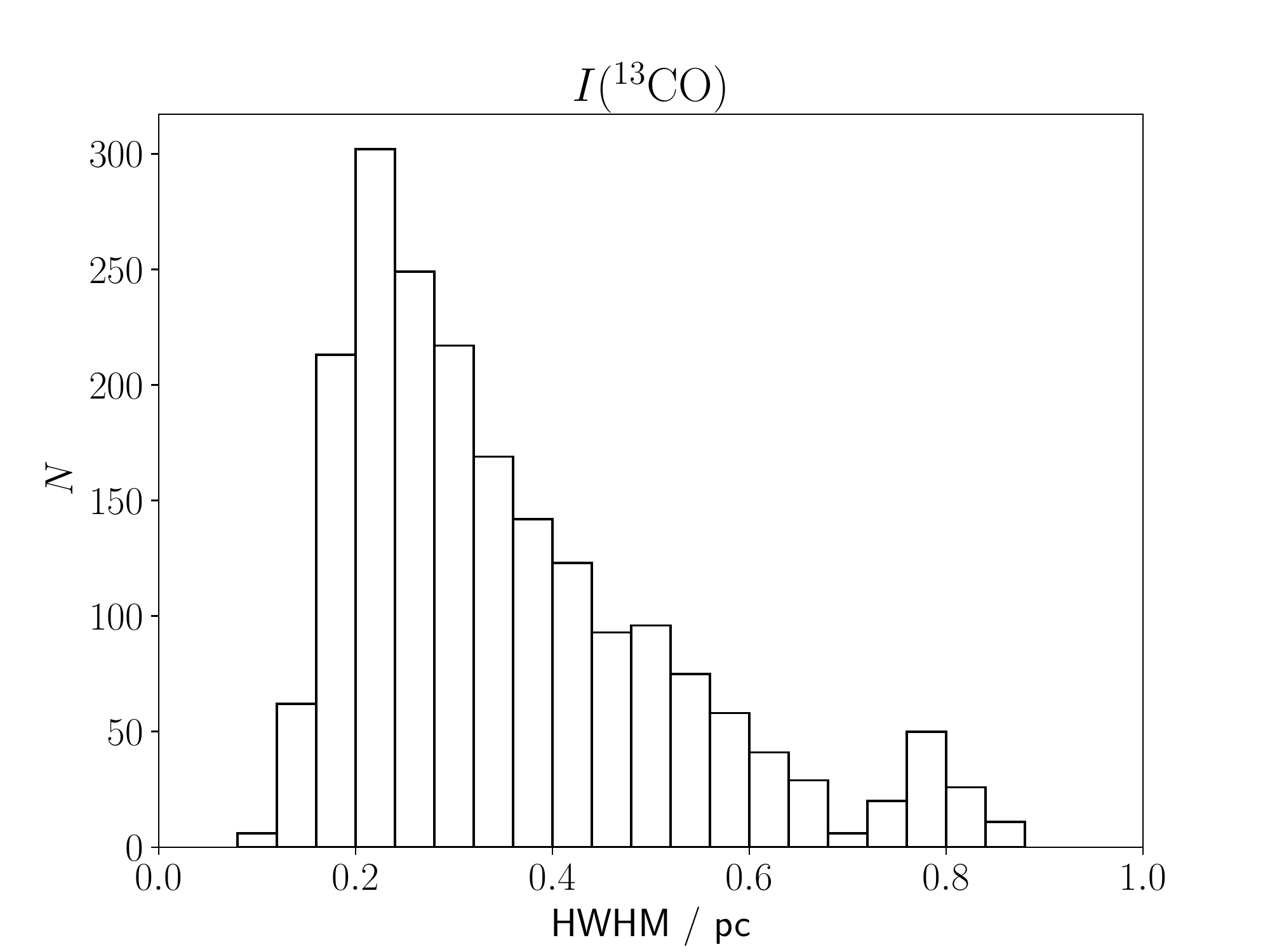}}\quad
  \subfigure{\includegraphics[width=\columnwidth]{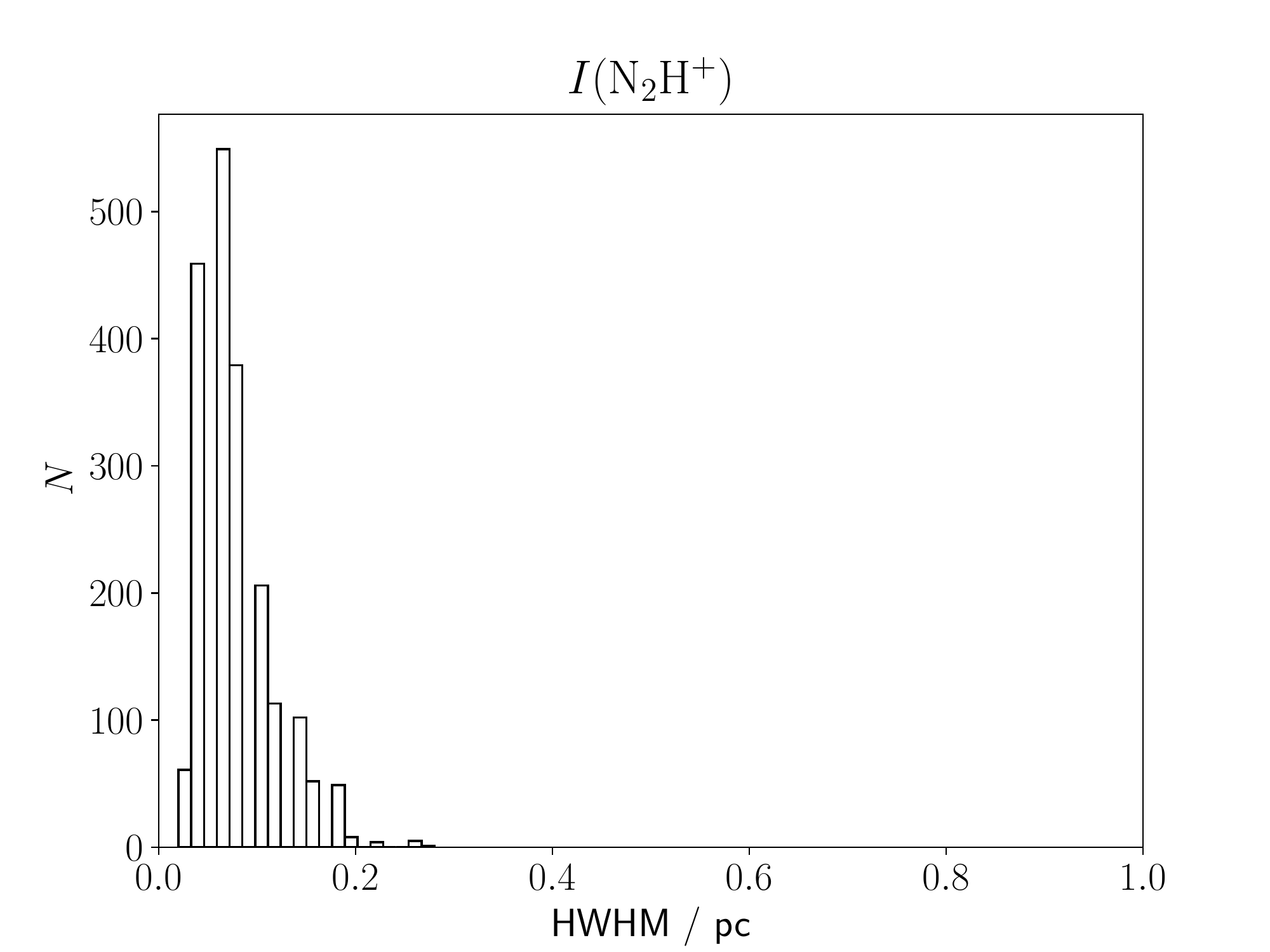}}
  \caption{Distributions of filament HWHMs for the $G\!=\!1$ case at $0.868\,{\rm Myr}$, derived from maps of the actual surface density ({\it top left}), the $850 \um$ dust-continuum intensity ({\it top right}), the $^{13}$CO (1-0) intensity ({\it bottom left}), and the N$_2$H$^+$ (1-0) intensity  ({\it bottom right}).}
  \label{fig:widths}
\end{figure*}

Filament widths are usually estimated by fitting an assumed functional form to the profile perpendicular to the filament spine \citep{arzoumanian2011,federrath2016,suri2019}, most commonly a Plummer-like profile,
\begin{eqnarray}\label{EQN:Plummer.1}
\Sigma(w)&=&\Sigma_{_{\rm O}}\,\left\{1\,+\,\left(\!\frac{w}{w_{_{\rm O}}}\!\right)^2\right\}^{\!-p/2}\,,
\end{eqnarray}
where $\Sigma_{_{\rm O}}$ is the value on the spine, $w_{_{\rm O}}$ is the scale-length close to the spine, and $p$ is the envelope exponent \citep{whitworth2001}. As the profiles in our simulations are not necessarily well represented by this (or any other) functional form, and in some instances are not even particularly symmetrical, we instead measure the widths by calculating the Half Width at Half Maximum (HWHM), separately on each side of the filament spine. Figure \ref{fig:widths} shows the distributions of HWHMs for the $G\!=\!1$ case, estimated on the basis of the actual surface density, and on the basis of the intensities of $850\,\mu{\rm m}$ dust-continuum emission, $^{13}$CO (1-0) line emission and N$_2$H$^+$ (1-0) line emission. The corresponding median widths, and the $16^{\rm th}$ and $84^{\rm th}$ percentiles, are presented in Table \ref{tab:widths}. In the fiducial case, $G\!=\!1$, the distributions of HWHMs for the actual surface density and for the $850\um$ intensity are quite broad, with median at $\sim 0.17\pc$.

\begin{figure*}
  \centering
  \subfigure{\includegraphics[width=\columnwidth]{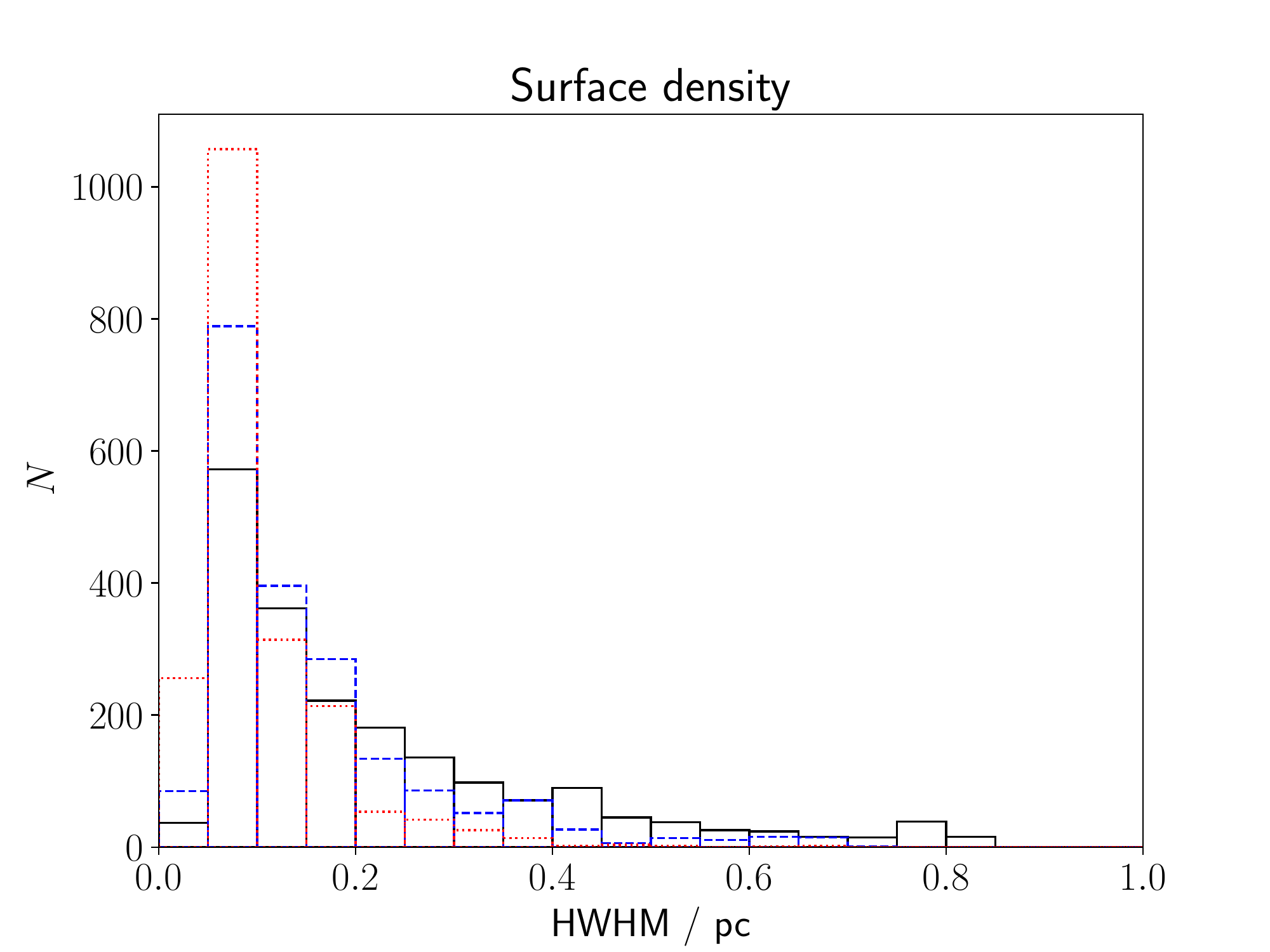}}\quad
  \subfigure{\includegraphics[width=\columnwidth]{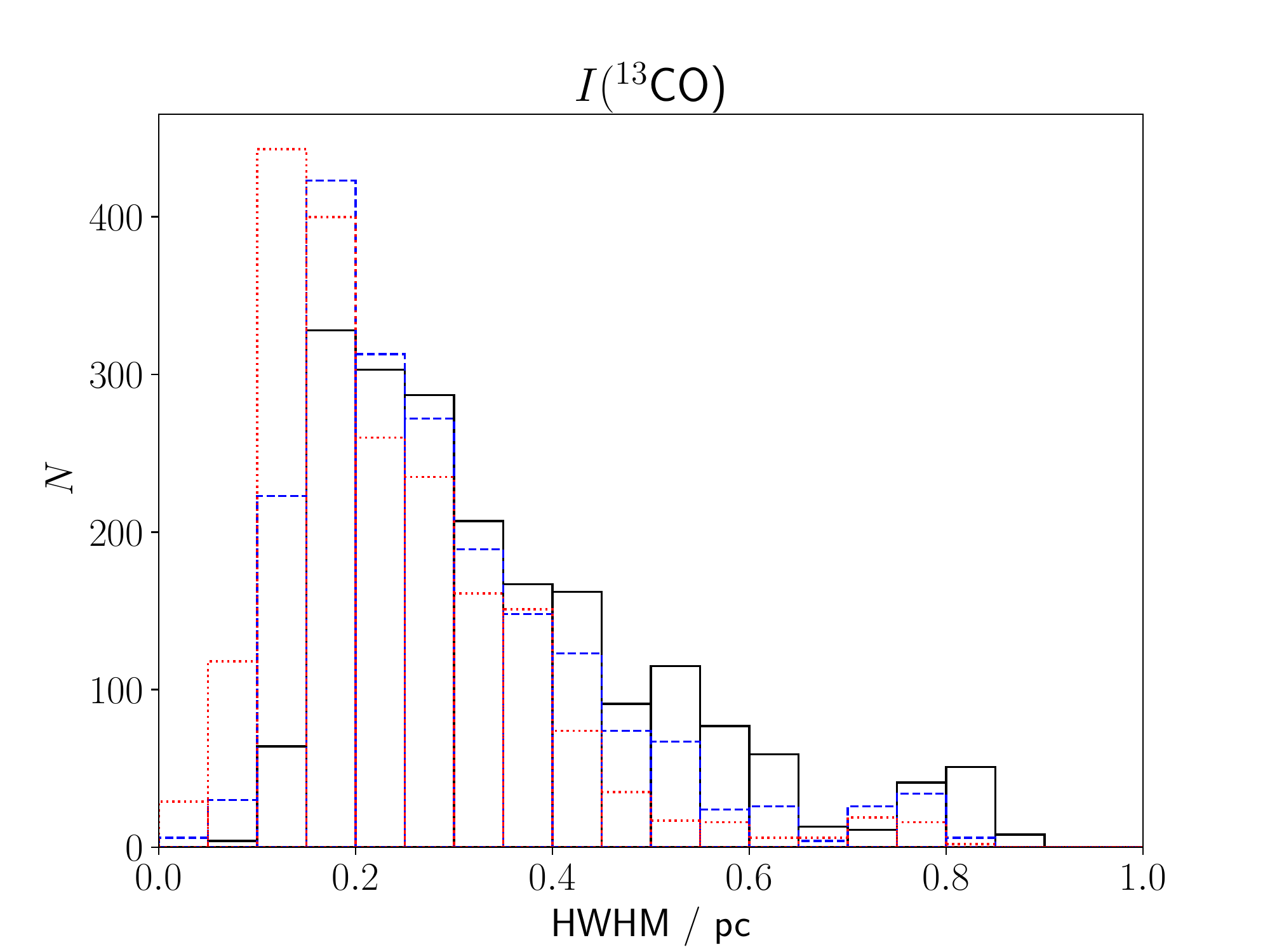}}
  \caption{{ Distributions of filament HWHMs for the $G\!=\!1$ case at $0.868\,{\rm Myr}$, derived from maps of the actual surface density ({\it left}) and the $^{13}$CO (1-0) intensity ({\it right}), with different levels of background subtraction. {\it Left:} no background subtraction (black solid line), $50 \msun \pc^{-2}$ background (blue dashed line) and $100 \msun \pc^{-2}$ background (red dotted line). {\it Right:} no background subtraction (black solid line), $1 \kel \kms$ background (blue dashed line) and $2 \kel \kms$ background (red dotted line).}}
  \label{fig:bgwidths}
\end{figure*}

The median HWHMs from our surface-density and dust-continuum maps are larger than the full-widths obtained both observationally \citep[e.g.][]{arzoumanian2011} and from previous simulations \citep[e.g.][]{federrath2016}, suggesting that our filaments are significantly wider. In fact, these authors refer to the FWHMs of fits to the filament profiles, rather than those of the profiles themselves. As can clearly be seen in several cases from \citet{arzoumanian2011,arzoumanian2019}, this often results in a FWHM smaller by a factor of up to a few, and for asymmetric, multiply peaked profiles such as the example shown in Figure \ref{fig:slice} this discrepancy can be even greater. If we instead estimate widths by fitting Plummer-like profiles to the region between the closest minima on either side of a peak, as in \citet{suri2019}, we obtain a much narrower distribution centred at smaller values, as found by previous authors. However, even with this procedure, the result depends sensitively on whether the envelope exponent for the Plummer-like fit, $p$ (see Equation \ref{EQN:Plummer.1}), is fixed or allowed to vary. Fixing $p$ often produces poor fits, while treating $p$ as a free parameter sometimes  produces unphysical values for the scale-length, $w_{_{\rm O}}$. Moreover, these issues tend to be more severe for line emission, because the intensity profiles are usually much less smooth. We shall therefore continue to use the HWHM to estimate filament widths. Whilst the exact values of filament widths depend on the procedure used to extract them, it is clear that there is a characteristic width of order $\sim 0.1 \pc$, and that it is well traced by the dust continuum.

HWHMs based on maps of molecular line emission are sometimes highly dependent on the strength of the ambient FUV radiation field, $G$, because this affects the abundances of the emitting molecules. The stronger  the ambient FUV radiation field, the more certain molecules become confined to the dense, well shielded inner parts of a filament, and consequently the narrower the filament appears to be when estimated using line radiation from such molecules. Conversely, the weaker the ambient FUV radiation field, the more extensively the molecule is distributed, and the wider the estimated HWHM. This effect is particularly notable for lines with a relatively low critical excitation density.

For example, in the fiducial case, $G\!=\!1$, HWHMs based on maps of $^{13}$CO (1-0) or C$^{18}$O (1-0) line emission are significantly larger than those based on maps of surface density or dust-continuum emission. Median values are  $\sim 0.30 \pc$ for $^{13}$CO (1-0) and $\sim 0.24\pc$ for C$^{18}$O (1-0), similar to the values seen by \citet{panopoulou2014} in the Taurus molecular cloud. However, if the FUV radiation field is increased to $G\ga 5$, HWHMs based on maps of $^{13}$CO (1-0) or C$^{18}$O (1-0) line emission become comparable with those based on maps of surface density or dust-continuum emission, because $^{13}$CO and C$^{18}$O then have very low abundance outside the dense inner regions of the filament. This may explain why \cite{suri2019} and \citet{orkisz2019} find that in Orion, where $G\ga 5$, HWHMs based on maps of C$^{18}$O (1-0) agree well with those based on dust emission.

HWHMs based on maps of line emission from molecules like N$_2$H$^+$, which have higher critical excitation densities and are therefore perceived as dense-gas tracers, are much lower than those from CO isotopologues, more tightly peaked, and on average even lower than HWHMs based on maps of surface density. This is largely due to the rapid decline in volume emissivity with decreasing volume density, leading to a minimal contribution from foreground and background diffuse gas; there may also be a small decrease in HWHM due to reduced abundance if the FUV radiation field is higher.

Although dense-gas tracers give filament widths $\la 0.1 \pc$, our values are still larger than those found by \citet{hacar2018}, $\lesssim 0.02 \pc$, using N$_2$H$^+$ observations. However, \citet{hacar2018} were measuring intertwined and approximately parallel `fibres' within a larger filamentary structure. Our simulations do not have sufficient resolution to distinguish such fibres, { with a typical smoothing length within the filaments of $0.01 \pc$,} and the effect of smearing them together would be to produce a single filament with width comparable to the estimates we obtain with $G\ga 1$.

{ A potential issue with our method of determining filament widths is that we assume all the emission/surface density is filamentary, without accounting for background contributions from the more diffuse material making up the cloud. In Figure \ref{fig:bgwidths}, we show the effect of subtracting increasingly stringent background values from the surface density and $^{13}$CO intensity on the filament width distributions, up to $100 \msun \pc^{-2}$ and $2 \kel \kms$ respectively. This reduces both the median and the range of the width distributions, but does not alter our conclusions. For a $2 \kel \kms$ background, the $^{13}$CO width distribution is still centred at significantly larger values ($0.2 \pc$) than that of the unsubtracted surface density, and for the highest value we consider the median surface density filament width is reduced to $0.08 \pc$. Similar results are obtained for other tracers: while the choice of background level can reduce the obtained characteristic width, the relative values between tracers are not greatly affected as long as consistent background levels are chosen. Even if a fairly high background level is assumed for some tracers while no subtraction is made for others, it is difficult to disrupt the hierarchy in Table \ref{tab:widths} of, for example, HWHM$_{\rm ^{13}CO} > $ HWHM$_\Sigma > $ HWHM$_{\rm HCN}$.}

\begin{figure}
  \centering
  \includegraphics[width=\columnwidth]{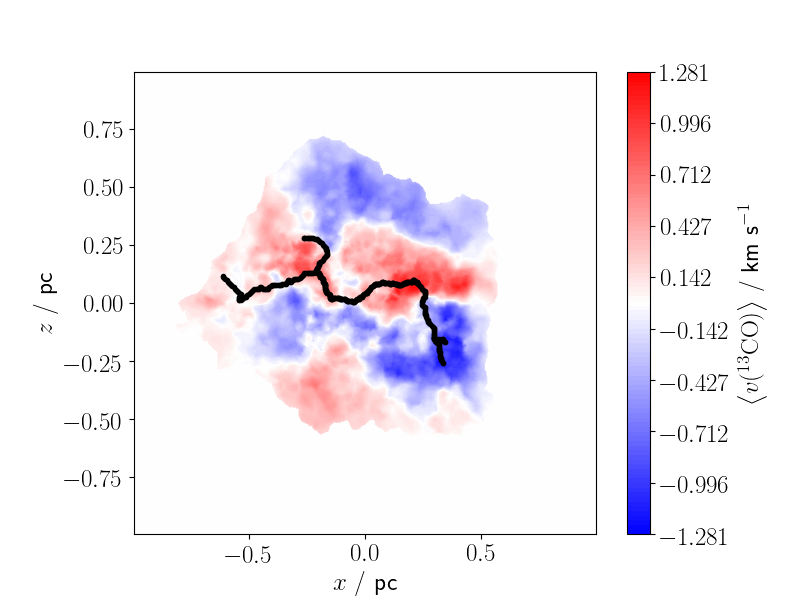}
  \caption{Intensity-weighted $^{13}$CO velocity map for the $G\!=\!1$ case at $0.868\,{\rm Myr}$, with the filament spine pixels coloured black.}
  \label{fig:velmap}
\end{figure}

\begin{figure}
  \centering
  \includegraphics[width=\columnwidth]{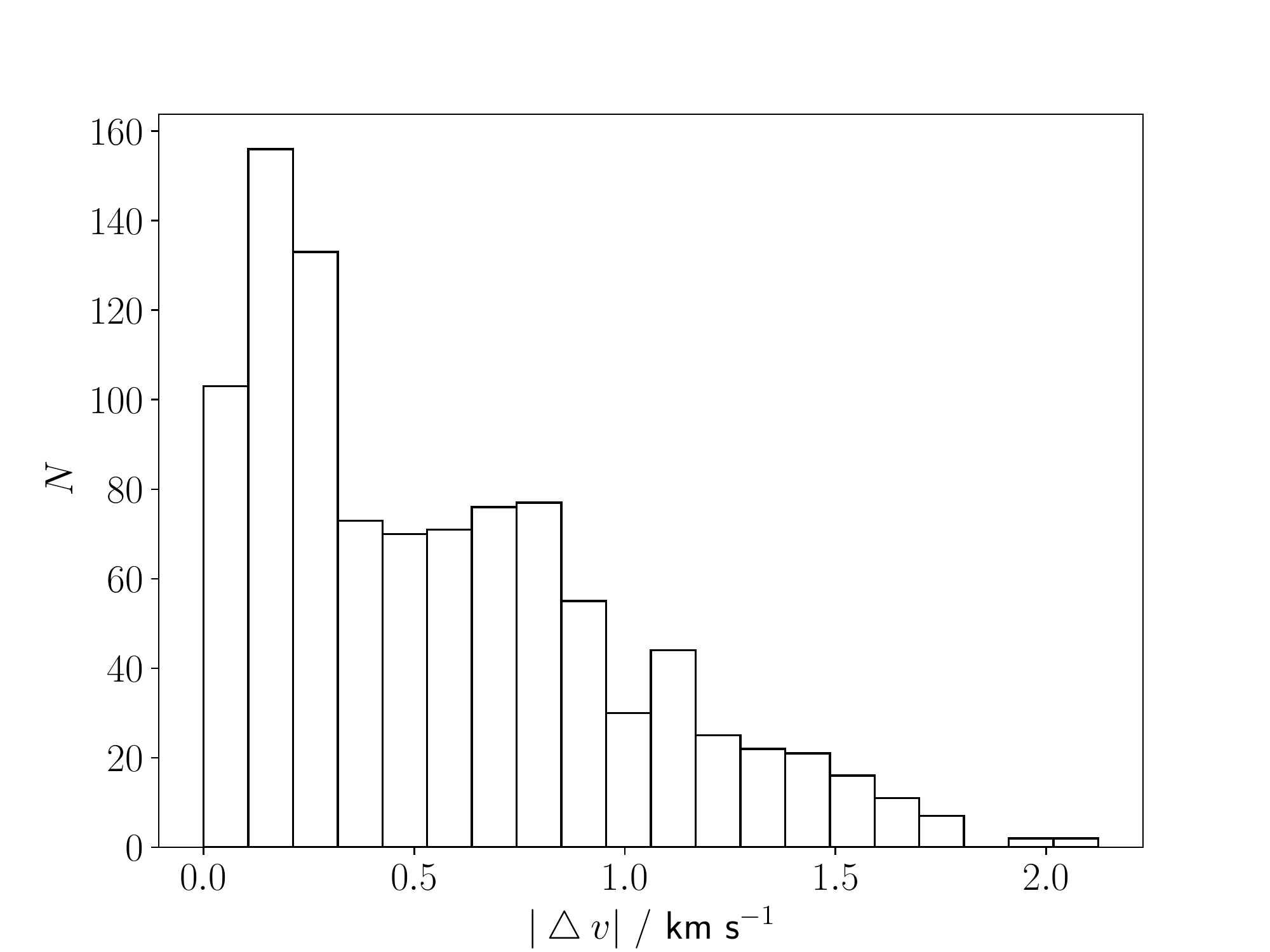}
  \caption{Distribution of $^{13}$CO radial velocity differences across the filament for the $G\!=\!1$ case at $0.868\,{\rm Myr}$.}
  \label{fig:velwidth}
\end{figure}

Figure \ref{fig:velmap} shows the first-moment map of the $^{13}$CO (1-0) intensity for the $G\!=\!1$ case at $0.868\,{\rm Myr}$, projected onto the $y\!=\!0$ plane (i.e. the $^{13}$CO intensity weighted mean $y$-component of velocity, $v_y$). The pixels defining the spine of the filament (as identified by DisPerSE) are black, and tend to be located between approaching and receding gas, i.e. where there is a marked radial velocity difference across the filament. Projections onto the $z\!=\!0$ and $x\!=\!0$ planes show a similar behaviour --- as do the maps for other molecular lines, although, due to the lower number of pixels with appreciable emission, this is less obvious for molecules like N$_2$H$^+$.

We estimate the radial velocity difference according to,
\begin{eqnarray}\label{EQN:Shear.1}
\Delta v_s\!&\!=\!&\!|v_{_{\rm R}}({\boldsymbol r}_s\!+\!\mbox{\sc hwhm}_s^+\hat{\boldsymbol n}_s)\,-\,v_{_{\rm R}}({\boldsymbol r}_s\!-\!\mbox{\sc hwhm}_s^-\hat{\boldsymbol n}_s)|\,,\hspace{0.7cm}
\end{eqnarray}
for each spine pixel. In Equation (\ref{EQN:Shear.1}), $v_{_{\rm R}}$ is the $^{13}$CO intensity weighted mean radial velocity, ${\boldsymbol r}_s$ is the position of pixel $s$ on the $y\!=\!0$ plane, $\hat{\boldsymbol n}_s$ is the unit normal to the filament spine at pixel $s$, {\sc hwhm}$_s^+$ is the {\sc hwhm} of the filament at pixel $s$ on the positive side (as defined by the unit normal, $\hat{\boldsymbol n}_s$), and {\sc hwhm}$_s^-$ is the {\sc hwhm} of the filament at pixel $s$ on the negative side. Figure \ref{fig:velwidth} shows the distribution of radial velocity differences for the $G\!=\!1$ case at $0.868\,{\rm Myr}$. The distribution is peaked at $\sim 0.20\kms\;(\sim c_{_{\rm s}})$, but with a substantial tail at higher values. The median is $0.48\kms$, and the $16^{\rm th}$ and $84^{\rm th}$ percentiles are $0.14$ and $1.0 \kms$ respectively. { Analysing the plane-of-sky velocities of simulated filaments, \citet{smith2016} reached similar conclusions - filaments tend to be moving supersonically with respect to the surrounding gas, as a result of forming at the convergence points of the turbulent velocity field.}

Position-velocity analysis is potentially a means of constraining the mechanisms regulating the formation and evolution of molecular clouds \citep[e.g.][]{haworth2015}. In the present context, the radial velocity difference across a filament may be a discriminating signature of filament formation by turbulence (as here). The data presented in \citet{ragan2012} and \citet{watkins2019} show some indication for velocity gradients perpendicular to the filament long axes, but on larger scales than those simulated here, and rotation would produce a similar effect. \citet{arzoumanian2018} find evidence for red- and blue-shifted $^{13}$CO emission on alternate sides of a $\sim 1 \pc$-long filament identified in C$^{18}$O, which they attributed to interaction between the filament and an extended sheet-like structure. This is more comparable to the situation in our simulations, being of a similar size and clearly indicating large-scale motions of less dense material, rather than rotation in the filament itself. Whether this phenomenon is widespread, and whether it can be produced by models other than the fragmentation of a turbulent cloud, is deserving of future study.

\begin{figure*}
  \centering
  \subfigure{\includegraphics[width=\columnwidth]{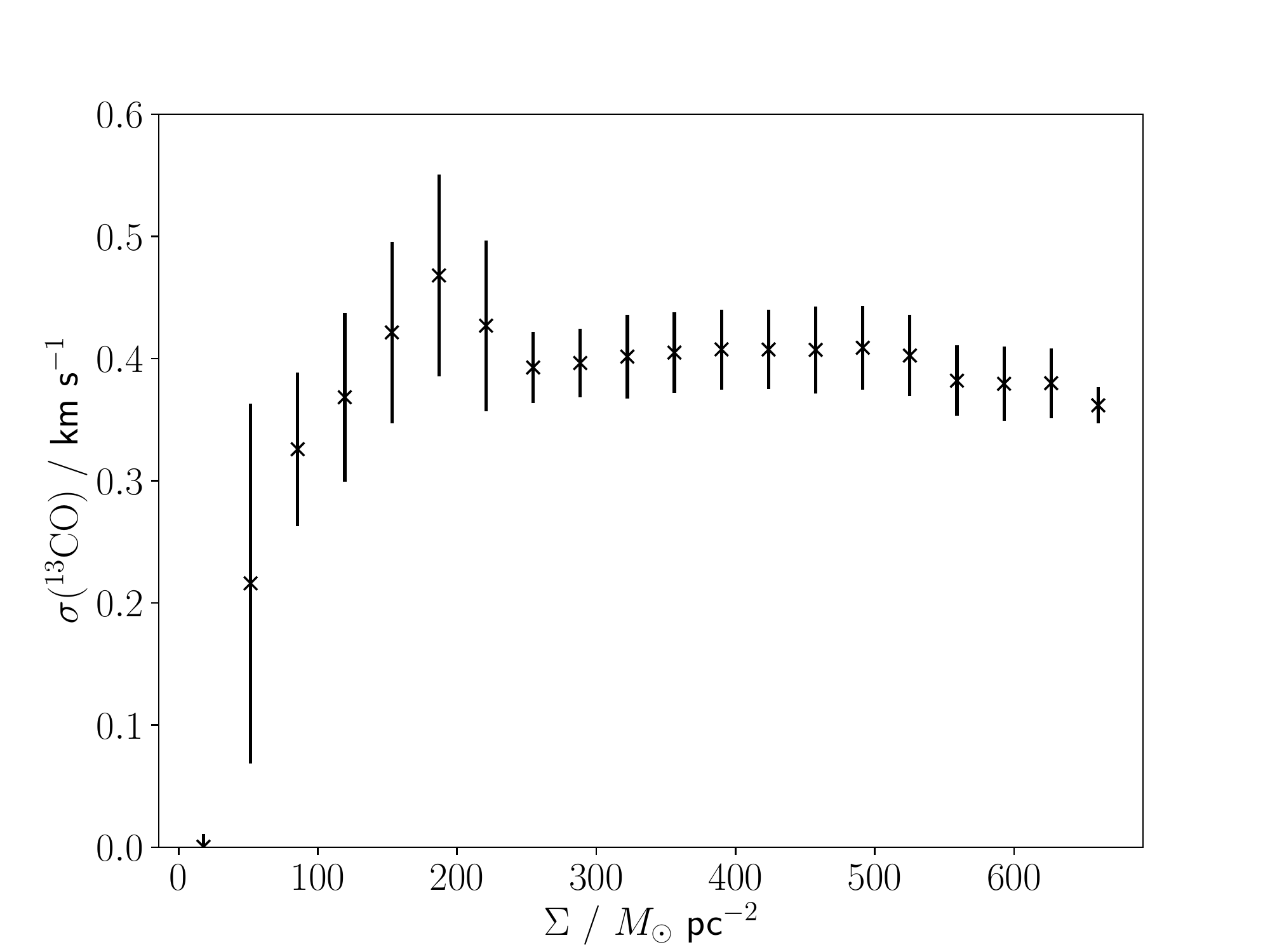}}\quad
  \subfigure{\includegraphics[width=\columnwidth]{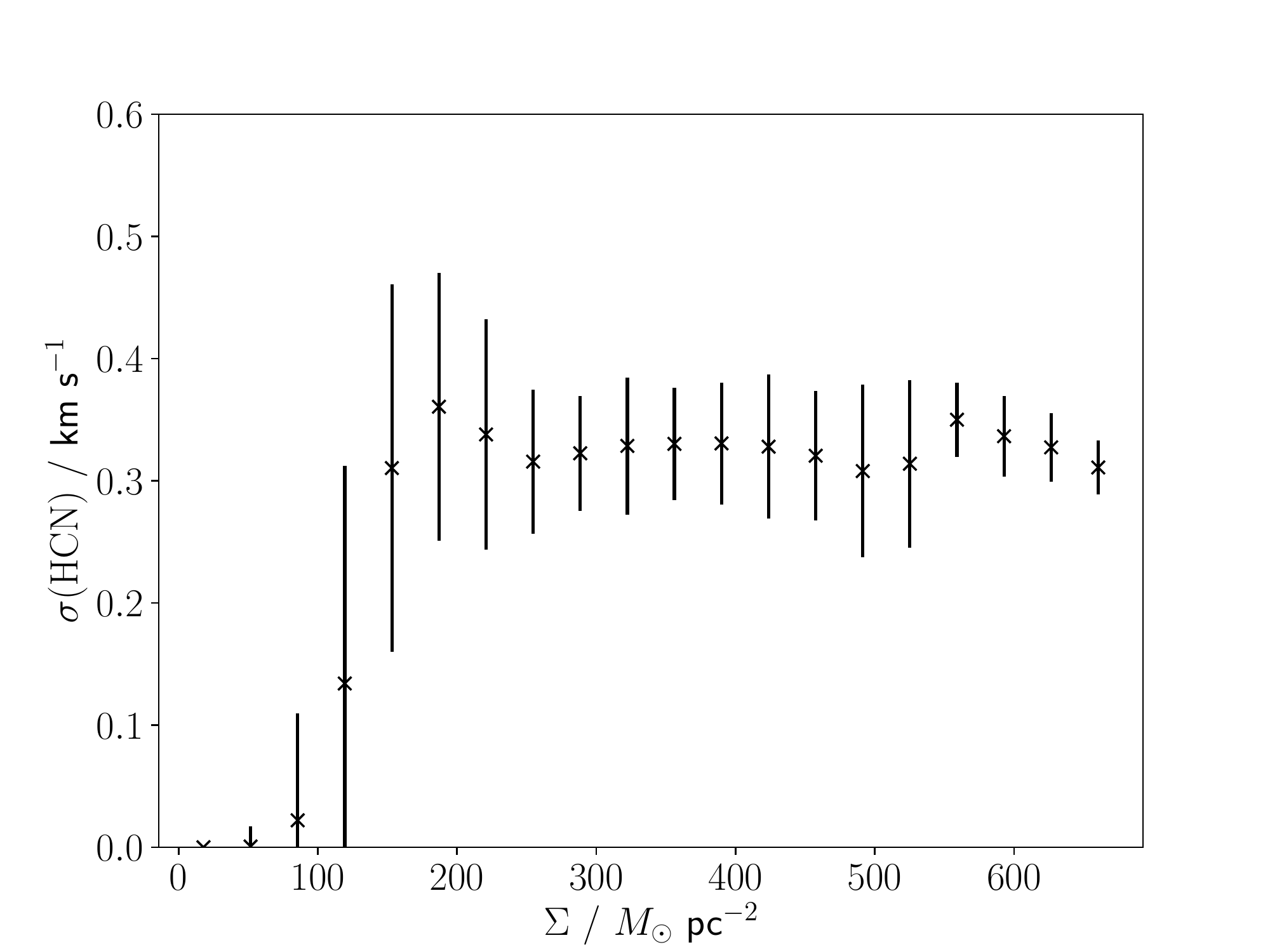}}
  \caption{Radial velocity dispersion versus surface density for the $G\!=\!1$ case at $0.868\,{\rm Myr}$, using $^{13}$CO ({\it left}) and HCN ({\it right}). The points give mean values, and the error bars give standard deviations.}
  \label{fig:turb}
\end{figure*}

Figure \ref{fig:turb} shows how the radial velocity dispersion varies with surface density for the $G\!=\!1$ case at $0.868\,{\rm Myr}$. The left plot uses the $^{13}$CO (1-0) line, and the right plot uses the HCN (1-0) line. At low surface density, $\Sigma<200\,{\rm M_{_\odot}\,pc^{-2}}$, in the outer parts of the cloud, the turbulence has decayed and the velocity dispersion is by this stage very low. At higher surface densities, $\Sigma\ga 200 \msun \pc^{-2}$, where the bulk of the gas is, the velocity dispersion is essentially independent of surface density, at  $0.3\,{\rm to}\,0.4 \kms$, i.e. mildly transsonic. There is a slight tendency for dense-gas tracers like HCN to display somewhat lower dispersions than CO isotopologues and NH$_3$. { Assuming, based on Figure \ref{fig:coldens}, that the structures traced by HCN and $^{13}$CO are approximately $0.5$ and $1 \pc$ in size respectively, the linewidth-size relation \citep{larson1981} predicts a velocity dispersion ratio of $\sim 0.75$ for an exponent of $0.4$, in good agreement with the actual values ($\sim 0.3$ and $0.4 \kms$ respectively).} These velocity dispersions are in good agreement with the observational results of \citet{arzoumanian2013} and \citet{hacar2018}.

\begin{figure*}
  \centering
  \subfigure{\includegraphics[width=\columnwidth]{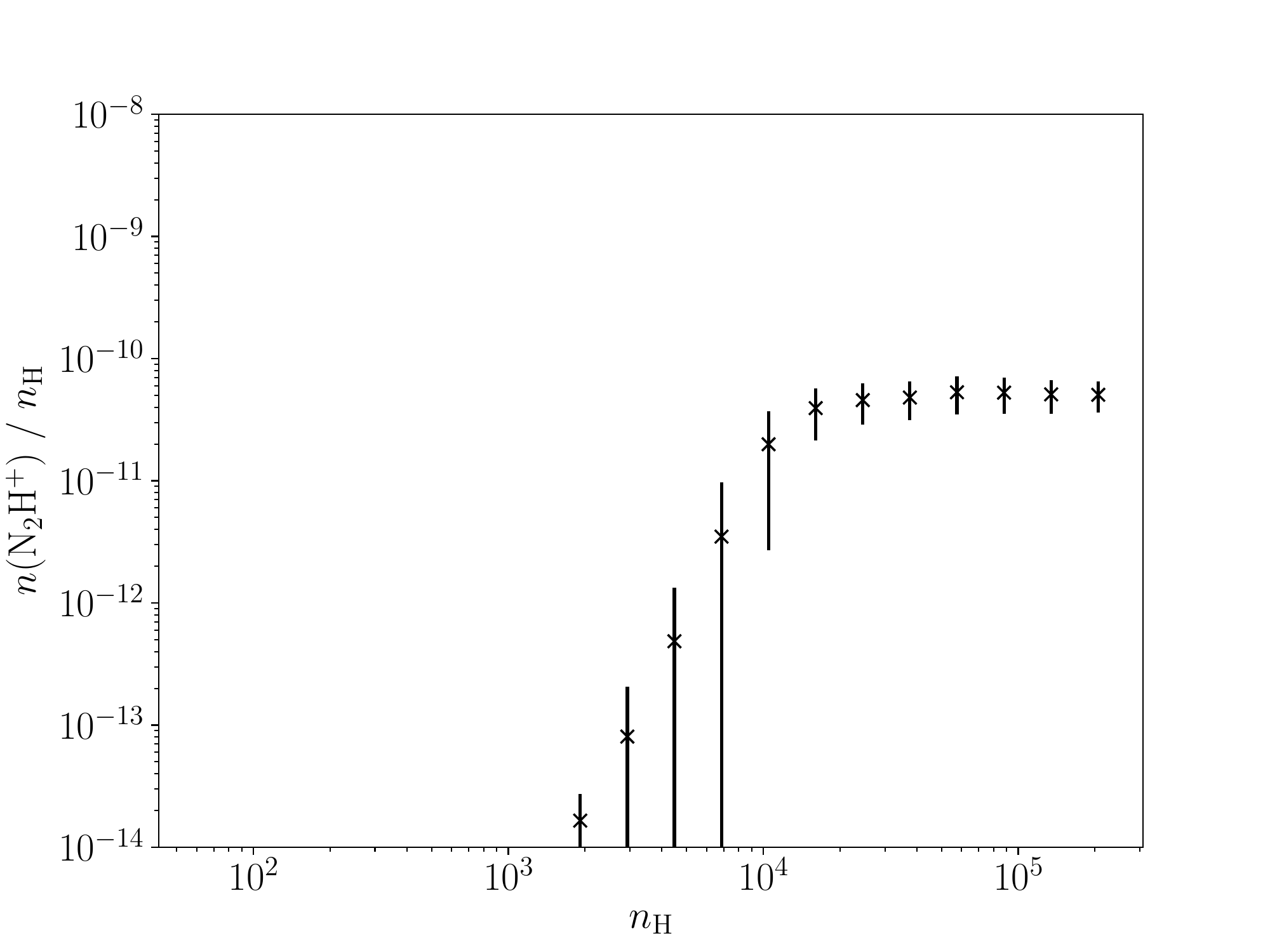}}\quad
  \subfigure{\includegraphics[width=\columnwidth]{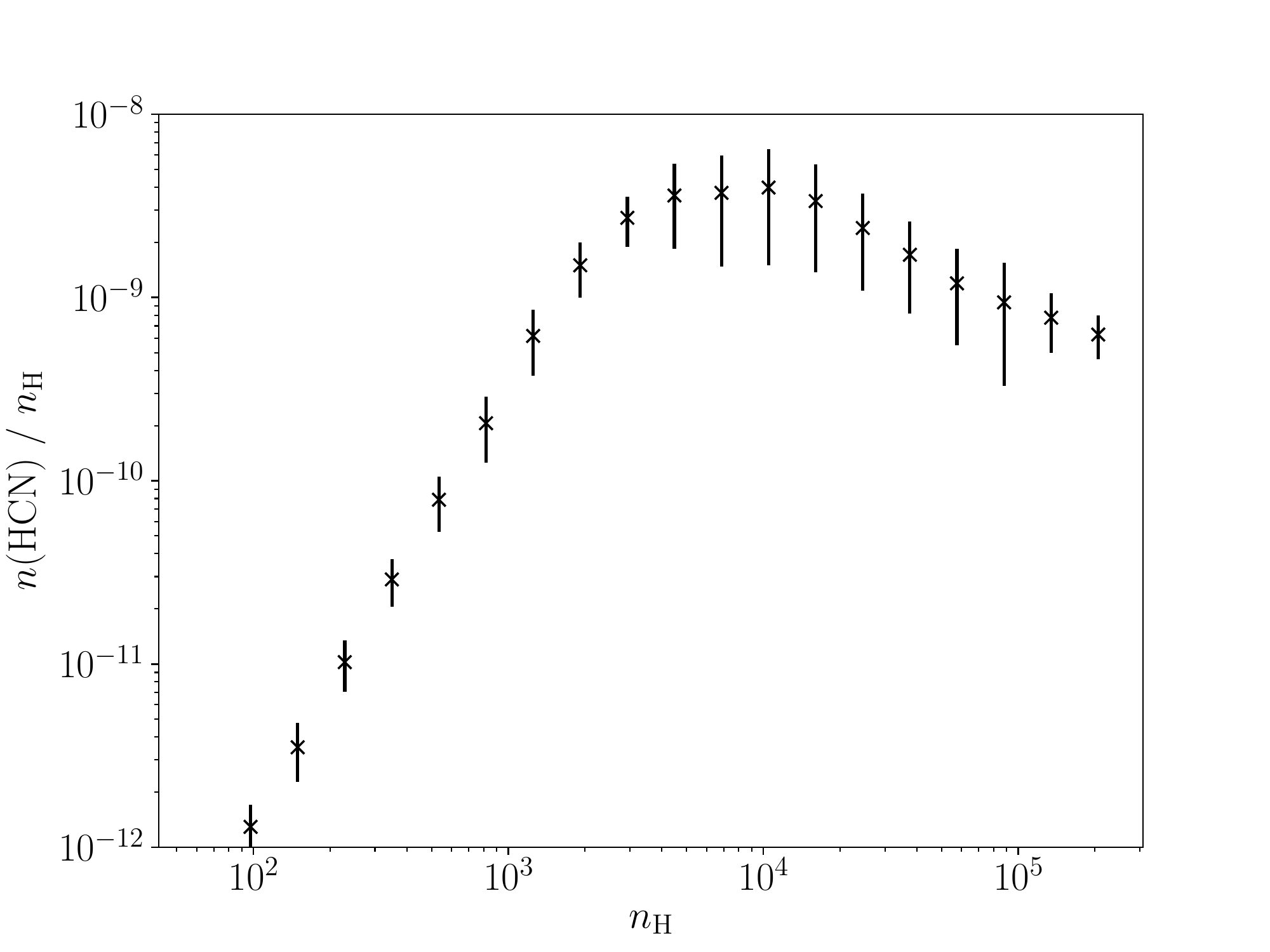}}
  \caption{Gas-phase molecular abundances versus the volume density of hydrogen in all forms ($n_{\rm H}$) for the $G\!=\!1$ case at $0.868\,{\rm Myr}$. {\it Left:} N$_2$H$^+$. {\it Right:} HCN. The points give mean values, and the error bars give standard deviations.}
  \label{fig:abun}
\end{figure*}

\begin{figure*}
  \centering
  \subfigure{\includegraphics[width=\columnwidth]{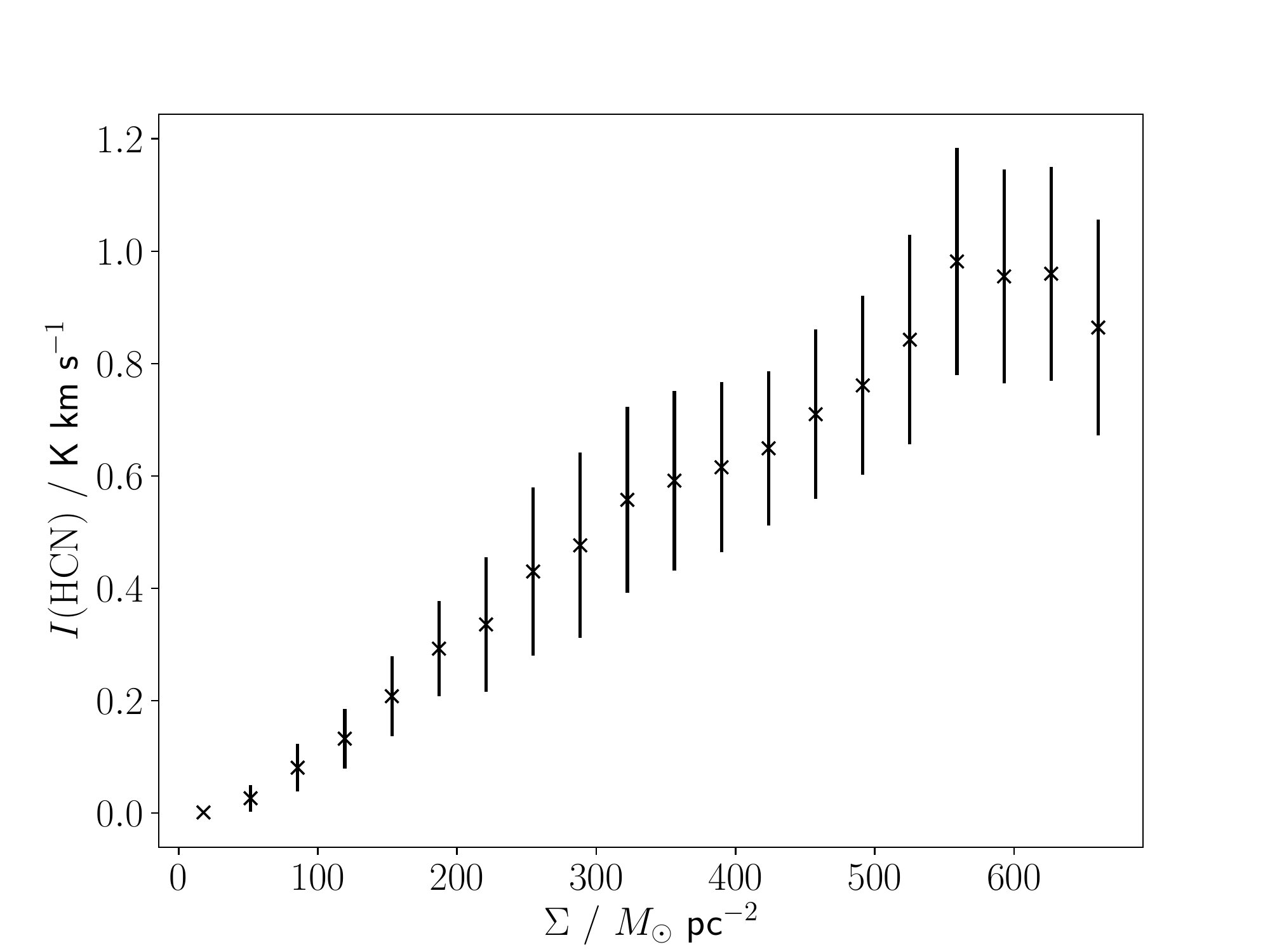}}\quad
  \subfigure{\includegraphics[width=\columnwidth]{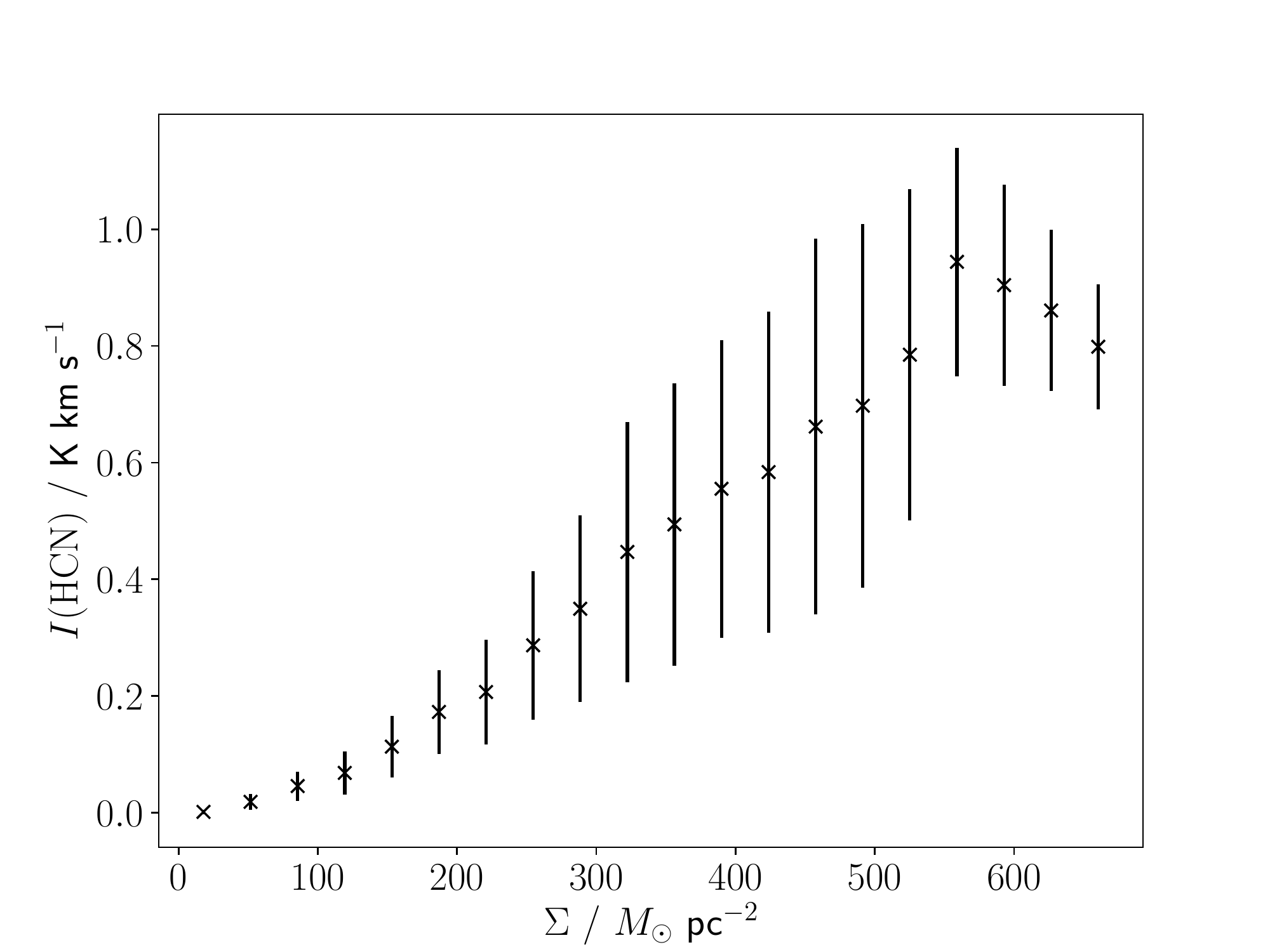}}
  \caption{{ The ratio of intensity to surface density on the $y\!=\!0$ plane for the $G\!=\!1$ case at $0.868\,{\rm Myr}$. {\it Left:} integrated line intensity of HCN $J=1-0$ emission {\it Right:} integrated line intensity of HCN $J=1-0$ emission assuming a constant abundance of $3 \times 10^{-9}$. The points give mean values, and the error bars give standard deviations.}}
  \label{fig:alphahcn}
\end{figure*}

\section{Discussion}

As with many previous studies, we find that the filaments which form in our simulations have a characteristic width of $\sim 0.1 \pc$. We have also shown that the far-IR dust emission is a reliable tracer of the gas surface density. Since we assume a constant universal dust temperature and constant universal dust properties, this is not surprising, since the dust emission is always optically thin ($\tau_{_{850\um}}\la 0.0002$). If we were to account for decreased dust heating in shielded regions, or the effect of growing ice mantles on the far-IR dust opacity, this might degrade the correlation between dust-continuum intensity and surface density, but it seems unlikely that these effects would be large enough to alter significantly the filament widths inferred from far-IR dust emission, at least for typical molecular cloud environments. 

Moreover, the filament widths derived using molecular-line intensities, for which we do account for the effect of environment (gas-phase abundance, optical depth, etc.), are generally in good agreement with those seen observationally. CO isotopologues tend to return larger widths, unless the ambient FUV radiation field is strong (as, for example, in Orion). Molecules like N$_2$H$^+$, which have high critical excitation densities, typically return widths similar to, or even lower than, those inferred from far-IR dust emission. While our treatment of the FUV radiation field is crude, altering the strength from zero to five times the local value does not qualitatively change our conclusions. Thus our results are broadly consistent with the universal characteristic filament width identified by \citet{arzoumanian2011,arzoumanian2019}.

Since we have calculated the molecular abundances self-consistently with the hydrodynamics, we can evaluate the accuracy of the abundance prescriptions used by previous authors. Figure \ref{fig:abun} shows how the  abundances of N$_2$H$^+$ and HCN vary with the volume density of hydrogen in all forms,
\begin{eqnarray}\label{EQN:nH}
n_{_{\rm H}}&=&2\,n_{_{\rm H_2}}\,+\,n_{_{\rm H^o}}\,+\,n_{_{\rm H^+}}\,+\,.\,.\,.\,,
\end{eqnarray}
where the additional unspecified terms on the righthand side of Equation (\ref{EQN:nH}) represent molecules (e.g. H$_2$O, NH$_3$, etc.) which are so rare that their contributions to the sum can safely be neglected. \citet{smith2012,smith2013} assume constant abundances for both N$_2$H$^+$ and HCN, and N$_2$H$^+$ does indeed have a roughly constant abundance at densities above $\sim 10^4 \pcc$, albeit somewhat smaller than the value adopted by \citet{smith2012}. However, the abundance of HCN is not even approximately constant over any significant density range; in particular, the abundance decreases by a factor of $\sim 5$ between $n_{_{\rm H}}\sim 10^4\pcc$ and $n_{_{\rm H}}\sim 10^5\pcc$, and probably continues to decrease at even higher densities. { As the HCN (1-0) transition is typically optically thick at these densities, the decrease in abundance may not greatly affect the resulting line intensities, but the large change in abundance around the effective excitation density of $8 \times 10^3 \pcc$ (the density required to produce a line intensity of $1 \kel \kms$; \citealt{shirley2015}) may have greater effects. We find that assuming a constant HCN abundance of $3 \times 10^{-9}$ \citep{smith2012} reduces the absolute value of, and increases the scatter in, the line intensity as a function of column density, shown in Figure \ref{fig:alphahcn}, although as this is only by a modest factor ($\sim 1.5$) it may not be of great importance. This, however, is not necessarily the case for all molecules, or for differing values of the radiation field and other input parameters which can significantly affect the gas-phase abundances. In any case, HCN emission does not become optically thick until a surface density of $\sim 550 \msun \pc^{-2}$, compared to around $200 \msun \pc^{-2}$ for $^{13}$CO. For a cloud depth of $1 \pc$, this corresponds to an average volume density of $<\nh> = 1.6 \times 10^4 \pcc$, two orders of magnitude below the critical density, and within a factor of a few of the effective excitation density from \citet{shirley2015}.} Recent work by \citet{evans2020} has found that the majority of HCN emission does not actually appear to originate from dense gas, corroborating this conclusion.

Whilst we have demonstrated that a simple physical model of an isolated turbulent cloud is consistent with the observed properties of molecular filaments, it remains to be seen whether more complex models of molecular clouds also have these properties. \citet{federrath2016} find that the inclusion of magnetic fields does not significantly affect the properties of filaments formed in their simulations. However, dynamically important magnetic fields can have a major effect on the abundances of key molecules, particularly in the densest regions \citep{tassis2012,priestley2018,priestley2019}. Other models of filament formation, such as cloud-cloud collisions \citep[e.g.][]{balfour2015}, will likely show similar chemical evolution, but the different kinematic structures involved could generate readily observable statistical differences in position-velocity maps based on molecular line emission. We intend to investigate these possibilities in future work.

\section{Conclusions}

We have post-processed hydrodynamical simulations of an isolated turbulent isothermal molecular cloud, using  a time-dependent chemical network and a radiative transfer model. Our main conclusions are as follows:
\begin{itemize}
\item{The resulting filamentary structures have a characteristic width of $\,\sim\!0.1\pc$, which is clearly defined both on maps of the actual surface density, and on maps of $850\um$ dust-continuum emission.}
\item{Filament widths determined from CO isotopologues are several times larger than $0.1\pc$, because, at surface densities $\Sigma\ga 250\,{\rm M_{_\odot}\,pc^{-2}}$, the line intensity is poorly correlated with the surface density.}
\item{Molecules used to trace dense gas, such as N$_2$H$^+$ and HCN, typically give filament widths similar to, or smaller than, those measured directly from the surface density.}
\item{This is broadly consistent with current observational data, and suggests that the characteristic width identified by \citet{arzoumanian2011} is a real physical phenomenon.}
\item{Velocity dispersions within filaments are transsonic ($\sim 0.35(\pm0.05)\,\kms$) and roughly constant, up to the highest densities that our simulations reach.}
\item{We find evidence for significant radial velocity differences ($\sim 0.4\kms$) across filaments, and we suggest that this may be a useful discriminatory signature of turbulent filament formation.}
\item{We confirm that HCN may not be a reliable tracer of dense gas, because { its (1-0) transition becomes optically thick at surface densities well below those expected based on its critical density of $\sim 10^6 \pcc$.}}
\end{itemize}

\section*{Acknowledgements}

{ We are grateful to the referee for several constructive suggestions.} We thank Paul Clark, Liz Watkins, Ollie Lomax, Niall Jeffrey and James Wurster for their assistance with various aspects of this paper. FDP and APW acknowledge the support of a Consolidated Grant (ST/K00926/1) from the UK Science and Technology Facilities Council (STFC).

\section*{Data Availability}

The data underlying this article will be shared on request. The codes used are all publicly available, at: phantomsph.bitbucket.io ({\sc phantom}); uclchem.github.io ({\sc ucl\_chem}); github.com/lime-rt/lime ({\sc lime}).

\bibliographystyle{mnras}
\bibliography{turbfil}

\bsp	
\label{lastpage}
\end{document}